\documentclass[12pt,a4-paper]{article}

\usepackage[utf8]{inputenc}

\usepackage{amsmath,amssymb} 
\usepackage{accents}
\usepackage{xcolor}
\usepackage[bookmarks=true,hyperfigures=true]{hyperref}
\usepackage{graphicx}
\usepackage[bulletsep]{collref}
\usepackage{tensor}
\usepackage{shuffle}
\usepackage{lscape}
\usepackage{empheq}
\usepackage{feynmf}
\usepackage{slashed}
\usepackage{caption}
\usepackage{graphbox}
\usepackage{float}
\usepackage{mathtools}

\usepackage[numbers, sort&compress]{natbib} 

\allowdisplaybreaks

\usepackage[a4paper,text={173mm,216mm},centering]{geometry}

\usepackage[font=small,labelfont=bf,width=0.85\textwidth]{caption}

\newcommand{\bra}[2]{\left[#1 #2\right]}
\newcommand{\ket}[2]{\langle #1  #2\rangle}
\newcommand{\ketbra}[3]{\left< #1\vert #2 \vert #3\right]}
\newcommand{\mombraket}[2]{\left| #1 \right]^{\dot{#2}} \left< #1 \right|^{#2}} 

\newcommand{\bee}{\begin{eqnarray*}}
\newcommand{\eee}{\end{eqnarray*}}
\newcommand{\be}{\begin{eqnarray}}
\newcommand{\ee}{\end{eqnarray}}

\newcommand{\effequal}[0]{\stackrel{\mathclap{\normalfont\mbox{eff}}}{=}}

\newcommand{\YMS}{YM+$\phi^3$}
\newcommand{\theory}{EYM}
\newcommand{\gym}{g_{\text{YM}}}

\begin{document}

\begingroup\raggedleft\footnotesize\ttfamily
HU-EP-18/36
\vspace{15mm}
\endgroup

\begin{center}
{\LARGE\bfseries Positive helicity Einstein-Yang-Mills 
amplitudes from the Double Copy

\par}%

\vspace{15mm}

\begingroup\scshape\large 
Josua~Faller and Jan~Plefka
\endgroup
\vspace{5mm}

\textit{Institut f\"ur Physik und IRIS Adlershof, Humboldt-Universit\"at zu Berlin, \phantom{$^\S$}\\
  Zum Gro{\ss}en Windkanal 6, D-12489 Berlin, Germany} \\[0.25cm]
  
\bigskip
  

\vspace{8mm}


\textbf{Abstract}\vspace{5mm}\par
\begin{minipage}{14.7cm}
All positive helicity four-point gluon-graviton amplitudes in Einstein-Yang-Mills theory
coupled to a dilaton and axion field are computed at the leading
one-loop order using colour-kinematics duality. 
In particular, all relevant contributions in the gravitational
and  gauge coupling are established. 
This extends a previous generalized unitarity based computation beyond 
the leading terms in the gravitational coupling $\kappa$. The resulting purely rational expressions take very compact forms. The previously seen vanishing of the
 single-graviton-three-gluon amplitude at leading order in $\kappa$ is seen to be lifted at order $\kappa^{3}$.
 \end{minipage}\par

\end{center}
\setcounter{page}{0}
\thispagestyle{empty}
\newpage

\hypersetup{
colorlinks=true,
linktoc=page,
citecolor=Blue,
linkcolor=Blue,
urlcolor=Blue}


	\setcounter{tocdepth}{4}
	\hrule height 0.75pt
	\tableofcontents
	\vspace{0.8cm}
	\hrule height 0.75pt
	\vspace{1cm}
	
	\setcounter{tocdepth}{2}

	\newpage

\section{Introduction}

 Scattering amplitudes involving only positive helicity gluon and graviton states take a very special r\^ole in minimally coupled gauge and gravitational theories: At tree-level they always vanish due to an effective (or hidden) supersymmetric Ward identity \cite{Grisaru:1977px}
	\be
		\mathcal{M}_{n,m}^{\text{tree}}(1^{+}_{a_{1}},\ldots,n^{+}_{a_{n}};(n+1)^{++},\ldots , (n+m)^{++})=0\,.
	\ee
This holds true for an amplitude in an arbitrary matter coupled Einstein-Yang-Mills (EYM)  with
 $n$ gluons and $m$ gravitons and 
persists to all-loops in supersymmetric theories. In non-supersymmetric
theories, in particular in the ``pure'' Yang-Mills (YM) and Einstein gravity 
examples, the leading contribution to these amplitudes is at the one-loop 
order. The resulting one-loop pure gluon or pure graviton positive helicity amplitudes turn out to be remarkably simple rational functions of the kinematic  invariants and
spinor products. This is a mandatory property in order to have vanishing unitarity cuts in four 
dimensions for these amplitudes. In fact, these pure gluon or graviton one-loop amplitudes are identical to the ones one finds in
self-dual YM or self-dual gravity, respectively. Here all-multiplicity expressions exists for the pure gluon and pure graviton amplitudes of uniform helicities \cite{Bern:1993qk,Mahlon:1993fe,Bern:1998sv}. 

Computing scattering amplitudes involving graviton via Feynman diagrams, as they follow from an expansion of the metric tensor about a flat Minkowski background, is a daunting task, because of the sheer complexity and the infinite number of vertices involved. An important insight that arose from the study of these amplitudes, however, is that they are much simpler than expected and display an intimate connection to amplitudes in YM theory. This relation is known as the double copy, as gravitational amplitudes may be obtained by taking the product of two gauge theory quantities.
 
The earliest such connection is due to Kawai, Lewellen and Tye (KLT) who expressed gravitational tree-level amplitudes as sums over products of gauge theory amplitudes weighted by Mandelstam invariants, originating from a string theory analysis \cite{Kawai:1985xq}. A decade ago 
Bern, Carrasco and Johansson (BCJ) \cite{Bern:2008qj} showed that these relations may be understood elegantly in terms of a specific diagrammatic expansion of the amplitudes: If one represents the gauge theory amplitudes in a trivalent fashion such that all kinematic numerators are arranged to obey a Jacobi-like relation mirroring the property of the colour degrees of freedom, then
the numerator of a gravitational amplitude follows  by simply squaring the gauge theory kinematic numerator. 
This duality -- known as the colour kinematic duality (CKD) -- was proven at tree-level \cite{Bern:2010yg}
and conjectured to hold at loop-level as well \cite{Bern:2010ue}. Plenty of examples have been presented which confirm this conjecture and furnish the state of the art technique to compute highest loop orders in supergravities and beyond, see e.g. \cite{Bern:2011rj,BoucherVeronneau:2011qv,Carrasco:2011mn,Bern:2013yya,Mogull:2015adi,He:2015wgf, Johansson:2014zca,Johansson:2015oia, Carrasco:2012ca, Chiodaroli:2014xia, Chiodaroli:2015rdg, Chiodaroli:2015wal, Anastasiou:2016csv, Johansson:2017bfl, Johansson:2017srf, Anastasiou:2017nsz,Bern:2017ucb, Bern:2018jmv}. Generally it is difficult to find numerators that satisfy CKD at loop-level, therefore it has been shown in \cite{Bern:2017yxu} that a modified double copy can be used to construct gravity integrands where contact terms are included due to the violation of kinematical Jacobi identities.  Another one-loop generalization of the KLT-formula has been presented in \cite{He:2016mzd,He:2017spx}.
 
In this paper we focus on the positive helicity sector of amplitudes for scattering processes involving gravitons interacting with gluons described by Einstein-Yang-Mills theory coupled to a dilaton and axion 
	field (\theory),%
	\footnote{In this paper we call Einstein-Yang-Mills coupled to a dilaton and axion field EYM for short, it is defined in eq.~(\ref{eq:EYMLagrangian}). The theory without these scalars we term `pure EYM'.}
as it arises as the low energy limit of bosonic string theory.
	 The evaluation of tree-level $S$-matrix elements in pure EYM has made rapid progress in the recent years and may be considered as completely solved. The key insight was to relate colour-ordered amplitudes of pure EYM to YM theory via an expansion of the form
	\be
			\label{eq:TreeYMandEYM}
		\mathcal{M}^{\text{tree}}_{n,m} \left(1, \cdots , n; h_1, \cdots , h_m \right) = 
		\sum_{\alpha \in \text{Perm}(2, \cdots ,n-1; h_1,\cdots , h_m )}
		  \mathcal{N}\left(1, \alpha ,n \right) A^{\text{tree}}_{\text{YM}} \left(1, \alpha , n\right),
	\ee
which has been proposed by \cite{Stieberger:2016lng} for one graviton and $n$ gluons based on a string computation and proven by field theory techniques \cite{Nandan:2016pya}. Higher graviton extensions of this result were presented in \cite{Nandan:2016pya,delaCruz:2016gnm,Chiodaroli:2017ngp}. The generalization to the entire single-colour-trace and multi-colour-trace sector has been carried out in \cite{Teng:2017tbo} and \cite{Du:2017gnh} by giving an algorithm to construct the coefficients $ \mathcal{N}\left(1, \alpha ,n \right)$. The colour ordered amplitudes in YM theory on the other hand have been determined in \cite{Drummond:2008cr}, which implies the complete solution of the problem of EYM amplitudes
at tree-level. An extension to arbitrary matter-single graviton amplitudes
was recently established in \cite{Plefka:2018zwm}. 

	The next step has been to tackle the loop amplitudes in (pure) EYM.  In \cite{Nandan:2018ody} all four point amplitudes at one-loop with one particle species circulating in the loop have been calculated for up to one negative helicity state. To make the calculation tractable the authors used the two-particle cut method in $4-2\epsilon$ dimensions. Restricting to one particle species circulating the loop allowed them to use supersymmetric Ward–Takahashi identities to replace the virtual particles in the loop by complex scalars, since the all-plus and all-plus-but-one-minus helicity amplitudes vanish in supersymmetric and supergravitational theories \cite{Grisaru:1977px}. However, these identities do not hold for a mixed propagation of gravitons and gluons in the loop and therefore this technique is not applicable in the most general setting.

	In this work we shall partly fill this gap by computing in four dimensions all positive helicity amplitudes of \theory\, at four points to all orders in $\kappa$ at one-loop
	precision.%
			\footnote{The all negative helicity amplitudes are trivially given by charge conjugation.}
We now use the double copy method to generate the integrands of \theory\, by tensoring the two gauge theory integrands of YM and Yang-Mills coupled to a biadjoint scalar (\YMS) as was established in a series of papers \cite{Chiodaroli:2014xia,Chiodaroli:2015rdg,Chiodaroli:2015wal,Chiodaroli:2016jqw,Chiodaroli:2017ngp}
starting from matter coupled $\mathcal{N}=2$ supergravities. Focusing on the all-plus sector the calculation is greatly simplified since in
YM theory in this case the four point amplitude only receives contributions from the box integrals, which implies that this side of the double copy trivially satisfies the CKD. Therefore in the \YMS\, sector also a non duality respecting representation may be used to perform the double copy. The later
we generate through usual Feynman diagrammatics.
 
We can report on rather compact formulae, the summary of our results 
reads \footnote{We use spinor bracket notation in the conventions of \cite{Henn:2014yza} and the Mandelstam variables $S = \ket{1}{2} \bra{2}{1}$, $T = \ket{2}{3} \bra{3}{2}$ and $U = \ket{1}{3} \bra{3}{1}$.}: 
	\be
			\nonumber
		\mathcal{M}^{\text{1-loop}}(1^+_a, 2^+_b, 3^+_c, 4^+_d) & = & \frac{i}{(4 \pi)^2} \frac{\bra{1}{2} \bra{3}{4}}{\ket{1}{2} \ket{3}{4}}
		\Big( \Big[- \frac{4}{3} \, \gym^4 \, f^{a'ab'}f^{b'bc'}f^{c'cd'}f^{d'da'} 	
						\\
						\nonumber
			& &			
			- \frac{\kappa^2 \gym^2}{12}  
			\left(  4 f^{abe'}f^{e'cd} \left(U-T \right) +N S \; \delta^{ab} \delta^{cd}  \right)
						\\
						\nonumber
		&&+ \frac{\kappa^4}{960}
			\delta^{ab} \delta^{cd} \left( 40 \; T U - \left( 2 + N_g\right) S^2  \right) 
			 \Big] + \text{perm} \Big)	,	
				\\
					\label{eq:results}
		 \mathcal{M}^{\text{1-loop}}(1^+_a, 2^+_b,3^{+}_c,4^{++})  & = &
		 - \frac{\kappa^3 \gym}{(8\pi)^2}  \frac{f^{abc}}{\sqrt{2}} \frac{\bra{4}{1} \bra{4}{2} \bra{4}{3} \bra{1}{2}}{\ket{3}{4}},
			 	\\
			 		\nonumber
 		 \mathcal{M}^{\text{1-loop}}(1^+_a, 2^+_b,3^{++},4^{++}) &= & 
		 \frac{i}{(4\pi)^2} \delta^{ab} \frac{\bra{1}{2}^2\bra{3}{4}^2}{\ket{3}{4}^2}
		 \Big(  
		 	-\frac{\kappa^2 \gym^2}{24}   N 	
		 	+ \frac{\kappa^4}{1440}  S \left(2+N_g\right)
		 \Big),
		 		\\
		 			\nonumber
		\mathcal{M}^{\text{1-loop}}(1^{++}, 2^{++},3^{++},4^{++}) & = & 
		-\frac{i}{\left(4\pi\right)^{2}} \kappa^4
		\left(\frac{\bra{1}{2} \bra{3}{4}}{\ket{1}{2} \ket{3}{4}}\right)^2
		\frac{S^2 +T^2 + U^2}{1920}
		\left(2 + N_g \right)
	\ee
	
	The outline of our paper is the following: In section \ref{sec:Procedure} we present the general strategy of our approach. In section \ref{sec:DoubleCopy} the BCJ representation in terms of cubic vertices of an amplitude is reviewed and we explain how gravity integrands can be obtained from gauge theory integrands by using the double copy technique. After we have discussed the construction of the one-loop integrands of \theory, we explain in section \ref{sec:strategy} how to obtain the amplitude. The next part, section \ref{sec:Amplitudes}, is devoted to the evaluation of the amplitudes presented in equation \eqref{eq:results}. The paper is supplemented by three appendices. In \ref{sec:regularization} the four-dimensional-helicity regularization scheme we are using is presented and \ref{sec:LagrangianFMRules} contains additional information about the double copy of \theory\, as well as the Feynman rules for computing the one-loop integrands of \YMS. The third appendix contains all the integrands of \YMS\, which shall be used to determine the integrands of \theory\, for one particular ordering of the external states.

\section{Preliminaries}
	\label{sec:Procedure}
	
\subsection{Review on double copy construction}
	\label{sec:DoubleCopy}

	Any $L$-loop gauge theory amplitude with all particles in the adjoint representation of the gauge group SU($N$) can be written as
	\begin{eqnarray}
				\label{eq:GTamplitude}
		\mathcal{A}^{L-\text{loop}}_m = i^{L-1} g^{m-2+2L} \sum_{\mathcal{S}_m} \sum_{j\in \Gamma}
		\int \frac{d^{d L} l}{ (2\pi)^{dL} } \frac{1}{S_j} \frac{c_j n_j}{\prod_{\alpha_j} D_{\alpha_j}},
	\end{eqnarray}
which separates the amplitude into three parts:
	\begin{itemize}
		\item The colour dependence is encoded in the colour factors $c_j$ which are a chain of the adjoint generators $i f^{abc}$.%
	\footnote{This coincides with the conventions given in \cite{Chiodaroli:2017ngp} which differ from the conventions in \cite{Chiodaroli:2014xia}.}
	 They obey the Jacobi identity which for the four particle case may be sketched as  $c_s = c_t + c_u$.\footnote{$c_s := i f^{abe'} i f^{e'cd}$, $c_t:=- i f^{ade'} i f^{e'bc}$ and $c_u:= -i f^{ace'} i f^{e'db}$}.	Furthermore, the adjoint generators $i f^{abc}$ are antisymmetric in their indices which implies that the colour factors are antisymmetric under permutation: $c_i = -c_j$.
		\item The set of all reduced Feynman propagators $1/(p^2 - m^2)$ associated to the $j^{\text{th}}$ graph are denoted by the inverse of the product $\prod_{\alpha_j} D_{\alpha_j}$. 
		\item The numerators $n_j$ account for the remaining kinematical dependence of the amplitude. Note that factors of $\pm i$ in the Feynman propagators are also absorbed in $n_j$.
	\end{itemize}

 The second sum runs over all distinct, nonisomorphic, trivalent graphs $\Gamma$ and the first one over all $|\mathcal{S}_m |= m!$ permutations of the external legs. Any overcounting of the $j^{\text{th}}$ diagram is removed by the symmetry factor $S_j$. Note that by using the identity $1=D_{\alpha_j}/D_{\alpha_j}$ any graph in a diagrammatic expansion can be made trivalent formally.
 
	Representing the amplitude in the form given in \eqref{eq:GTamplitude} reveals the parallel treatment of colour degrees of freedom $c_j$ and kinematical degrees of freedom $n_j$ and is especially powerful if one arranges the kinematical numerators in such a way that they obey the same algebraic relations as the corresponding colour factors%
	\begin{eqnarray}
		\label{eq:CKD}
			\left.
				\begin{array}{ll}
		 c_s  = c_t + c_u 
		 	\\
		c_i = -c_j 
				\end{array}
			\right\}
			 \implies
			\left\{
				\begin{array}{@{}l@{}}
		 n_s= n_t + n_u
		 	\\
		n_i = -n_j
				\end{array}
			\right. .
	\end{eqnarray} 

 	It has been conjectured by Bern, Carrasco and Johanson (BCJ) \cite{Bern:2008qj, Bern:2010ue} and shown at tree-level $(L=0)$ in refs.~\cite{Bern:2010yg,Tye:2010dd,Mafra:2009bz,Mafra:2010gj, BjerrumBohr:2009rd, Mafra:2011kj} that it is always possible to arrange all the numerators $n_i$ of a diagram in such a way that they obey \eqref{eq:CKD}.
 
 It is a striking feature of this colour-kinematics duality (CKD) that integrands for gravity amplitudes can be easily constructed from integrands of gauge theories if at least one set of the gauge theory numerators $n_j$ or $\tilde{n}_j$ satisfies \eqref{eq:CKD}:
	\begin{eqnarray}
			\label{eq:DCgravity}
		\mathcal{M}^{L-\text{loop}}_m = i^{L-1} \left(\frac{\kappa}{4}\right)^{m-2+2L} \sum_{\mathcal{S}_m} \sum_{j\in \Gamma_g}
		\int \frac{d^{d L} l}{ (2\pi)^{dL} } \frac{1}{S_j} \frac{\tilde{n}_j n_j}{\prod_{\alpha_j} D_{\alpha_j}}.
	\end{eqnarray}
The set $\Gamma_g$ includes all pairs of trivalent graphs from both gauge theories with numerators $n_j$ and $\tilde{n}_j$, respectively.
	This feature of obtaining gravity amplitude integrands has been proven at tree-level for pure gravity using BCFW recursion relations \cite{Britto:2005fq}
in four dimensions \cite{Bern:2010yg}. Furthermore, gauge invariance of both gauge theory amplitudes implies invariance of the corresponding double copied amplitude under linearized diffeomorphisms, i.e.~\eqref{eq:DCgravity} is an amplitude of \emph{some} gravity theory \cite{Chiodaroli:2017ngp}. The unitarity method can extend this feature to loop-level by reducing amplitudes containing loops to tree-level amplitudes and demanding that CKD holds for all cuts. However, a general proof is still missing 
\cite{Bern:2011qt}\footnote{We note that the conjectured formula  (\ref{eq:DCgravity})
apparently fails at five loop for maximal
 supergravity if one requests the numerator factors to satisfy
 the Jacobi identity. Such a representation could not be found for the maximal super-Yang-Mills
at five loops. Instead a generalized BCJ relation~\cite{BjerrumBohr:2010zs,Bern:2017ucb} is
required.}. This double copy (DC) construction can be viewed as a generalization of the Kawai-Lewellen-Tye relations \cite{Kawai:1985xq} and a loop generalization thereof.

	Since gauge theory integrands are much easier to calculate than gravity integrands, the DC prescription gives a powerful tool to build gravity integrands. In particular, it is enough to construct a certain number of master integrands, because all other integrands are determined by relation \eqref{eq:CKD}. Diagrammatically this can be sketched as
	\be
			\label{eq:PicCKD}
		\begin{minipage}{3.5cm}
			\includegraphics[scale=0.5]{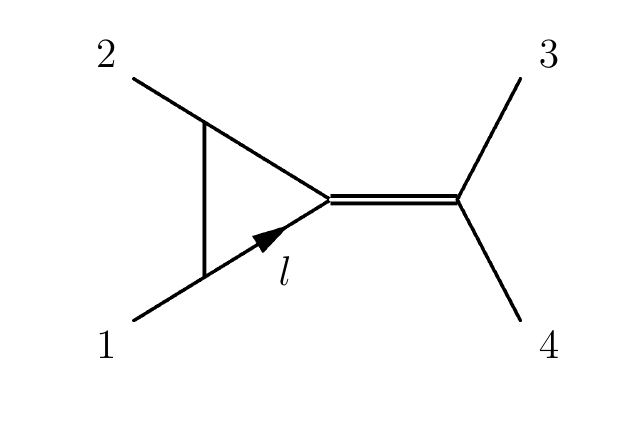}
		\end{minipage} 
		=  
		\begin{minipage}{3.5cm}
			\includegraphics[scale=0.5]{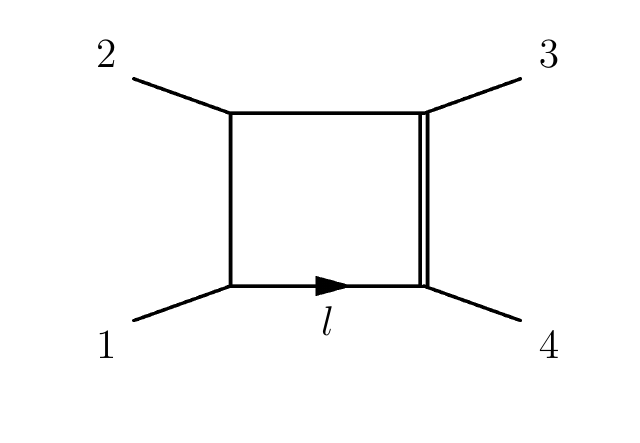}
		\end{minipage} 
		-		
		\begin{minipage}{3.5cm}
			\includegraphics[scale=0.5]{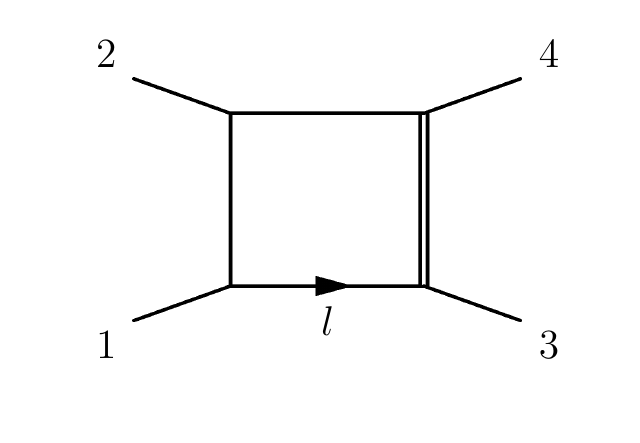}
		\end{minipage} 
	\ee

	Applying the same reasoning the bubble graphs may be reexpressed to triangle graphs. Therefore the master numerators for four points at one-loop are given by the box diagrams since all other diagrams can be obtained by applying the kinematical Jacobi identity \eqref{eq:CKD}. 
	\footnote{Here we do not consider the bubble-on-external-leg or tadpole diagrams that vanish in dimensional regularization after integration.}

\subsection{The double copy procedure for Einstein-Yang-Mills}
	\label{sec:strategy}

	In this section we outline the methods we use to obtain the amplitudes presented in \eqref{eq:results}. The strategy is to write the amplitude in the form  \eqref{eq:DCgravity}, for which we have to determine the numerators of pure Yang-Mills (YM) and Yang-Mills coupled to a biadjoint scalar (\YMS). Afterwards we perform the integrals by decomposing the tensor integral into scalar Feynman integrals and evaluating these in  four dimensions.
	
	We shall start our discussion by defining YM theory and \YMS\, theory through the Lagrangians 
	\be
				\label{eq:Lagr}
		\mathcal{L}^{\text{YM}} & = & -\frac{1}{4} F^{\mu \nu, a} F^a_{\mu \nu},
			\\
			\nonumber
		\mathcal{L}^{\text{\YMS}} & = & 
		\mathcal{L}^{\text{YM}} + \frac{1}{2} \left(D_{\mu} \phi^A \right)^a \left(D^{\mu} \phi^A \right)^a
		-\frac{g^2}{4}f^{abe} f^{ecd} \phi^{Aa} \phi^{Bb} \phi^{Ac} \phi^{Bd}
			\\		
			\label{eq:YMSCLagrangaian}
		&& + \frac{1}{3!}\lambda g F^{ABC} f^{abc} \phi^{Aa} \phi^{Bb} \phi^{Cc}
		\ee
		with the definitions
			\be
		F^c_{\mu \nu} & = & \partial_{\mu} A_{\nu}^c - \partial_{\nu} A^c_{\mu} + g f^{abc} A_{\mu}^a A^b_{\nu},
			\\
			\nonumber
		\left(D_{\mu} \phi^{A}\right)^a & = & \partial_{\mu} \phi^{Aa} + g f^{abc} A_{\mu}^b \phi^{Ac}.
	\ee
	We have chosen the normalization Tr$\left(T^a T^b\right)=\delta^{ab}$ for the Hermitian (fundamental) generators $T^a$ of the gauge group SU$(N)$, which implies that the structure constant is given by $f^{abc} = - \frac{i}{\sqrt{2}}\text{Tr}\left(T^a \left[T^b,T^c\right]\right)$. For the global gauge group with structure constants $F^{ABC}$
	it is simply demanded that they shall obey the Jacobi identity and that they are antisymmetric in all of their indices $A,B,C,\ldots$.%
		\footnote{After double copying we promote the global gauge group to the local gauge group SU$(N)$ of \theory.}
Dimensional analysis reveals that the coupling constant $\lambda$ is of mass dimensions one.

	The EYM Lagrangian emerging from the double copy of  \eqref{eq:Lagr} and \eqref{eq:YMSCLagrangaian} reads explicitly \cite{Johansson:2018ues}
	\be
			\label{eq:EYMLagrangian}
		\mathcal{L}^{\mathcal{N}=0+\text{YM}} = 
		\frac{\sqrt{-G}}{\kappa^2} 
		\left(
			-2\;R + \partial_{\mu} \varphi \partial^{\mu} \varphi +e^{2\varphi} \partial_{\mu} \chi \partial^{\mu} \chi \right)
			- \frac{\sqrt{-G}}{4} \left(
				e^{-\varphi} F^a_{\mu \nu} F^{\mu \nu, a} + i \chi  F^a_{\mu \nu} \tilde{F}^{\mu \nu, a}
			\right)\, ,
	\ee
where $G$ is the determinant of the metric
	$G_{\mu\nu}$, the scalar curvature is encoded in the Ricci tensor $R$ and $\tilde{F}^a_{\mu \nu} = \frac{i}{2} \sqrt{-G} \, \epsilon_{\mu \nu \varrho \sigma} F^{\varrho \sigma,a}$ represents the dual field strength tensor. The scalars $\varphi$ and $\chi$ are the dilaton and axion, respectively.%
		\footnote{In four dimensions the antisymmetric B-field, which naturally appears in $\mathcal{N}=0$ supergravity, can be replaced by a scalar using a duality transformation. Details can be found e.g. in \cite{Schwarz:1992tn}.}%
	 The authors of ref.~\cite{Johansson:2018ues} confirmed at tree level that the DC procedure of the theories \eqref{eq:Lagr} and \eqref{eq:YMSCLagrangaian} give the same amplitudes as the one derived from \eqref{eq:EYMLagrangian}. Clearly the dilaton as well as the axion interact with the gauge fields. To obtain the pure EYM result one has to remove the contributions originating from the two scalars. In \cite{Johansson:2014zca} it has been shown that for $\mathcal{N}=0$ supergravity the axion and dilaton can be subtracted by a DC of fields in the fundamental representations with opposite statistics.
	
	After discussing the relevant Lagrangians of the theories, we start with the DC procedure. The first step is to determine both numerators $n^{\text{YM}}$ and $n^{\text{\YMS}}$ which can be extracted from the corresponding one-loop integrands of pure YM and \YMS. It turns out that the four point YM numerator $n_{1^+ 2^+ 3^+ 4^+}^{\text{YM}}$ at one-loop for an all-plus helicity amplitude is very simple because it is exclusively given by graphs with box topology.%
	\footnote{This property follows from maximal cuts \cite{Britto:2004nc}, because only box type diagrams of YM theory are consistent with its analytic structure and all other diagrams shall cancel. }
	Let us quickly review this property. The corresponding YM amplitude reads in the colour trace basis
	\be
				\label{eq:YMampl}
		\mathcal{A}^{1-\text{loop}}_4 \left(1^+_{a_1},2^+_{a_2},3^+_{a_3},4^+_{a_4} \right) & = &
		\sum_{\sigma \in \mathcal{S}_4/\mathbb{Z}_4} N \; \text{Tr}\left(T^{a_{\sigma(1)}}\cdots T^{a_{\sigma(4)}} \right) 
		A_{4;1} \left(\sigma(1^+),\cdots , \sigma(4^+) \right)
				\\
				\nonumber
		& & + \sum_{\sigma \in \mathcal{S}_4/ S_{4;3}} \text{Tr}\left(T^{a_{\sigma(1)}} T^{a_{\sigma(2)}} \right)  \text{Tr}\left(T^{a_{\sigma(3)}} T^{a_{\sigma(4)}} \right) 
		A_{4;3} \left(\sigma(1^+),\cdots , \sigma(4^+) \right),
	\ee
where $\mathcal{S}_4$ is the permutation group which is quotiented by the subgroups $\mathbb{Z}_4$ and $S_{4;3}$ that leave the single and double traces invariant, respectively. $N$ is the degree of the Lie group SU$(N)$. In this case the primitive amplitude $A_{4;1}$ is related to the partial amplitude $ A_{4;3}$ by $ A_{4;3} = 6A_{4;1}$ \cite{Bern:1990ux}.
In \cite{Bern:1991aq, Bern:1995db,Bern:1996ja,Brandhuber:2005jw} $A_{4;1}$ has been calculated to all orders in $\epsilon$. Its Veltman-Passarino \cite{Passarino:1978jh} reduced form reads	
	\bee
		A_{4;1}(1^+,2^+,3^+,4^+) = \frac{2 i}{\left( 4\pi \right)^{2-\epsilon}} 
		\frac{\bra{1}{2} \bra{3}{4}}{\ket{1}{2} \ket{3}{4}} I_4\left[\mu^4;S,T\right],
	\eee	
	where the scalar-box integral $I_4\left[\mu^4;S,T\right]$ is defined in the appendix \ref{sec:regularization} and $\bra{i}{j}, \ket{i}{j}$ are the helicity spinor brackets of the external momenta $p_i$, $p_j$. The helicity spinor representation of these read $p^{\dot{\alpha} \alpha} = \mombraket{i}{\alpha}$ (see \cite{Dixon:1996wi,Elvang:2013cua,Henn:2014yza} for reviews). $\mu$ is the fictitious mass of the propagating complex scalar field in the loop which needs to be integrated over \cite{Bern:1995db} in order to emulate $4-2\epsilon$ dimensions.%
	\footnote{For more details on this regularization scheme see section \ref{sec:regularization}.}
Thus the corresponding numerator of the partial amplitude is
	\be
		a_{1^+ 2^+ 3^+ 4^+}^{\text{YM}} = 2 \mu^4 \frac{\bra{1}{2} \bra{3}{4}}{\ket{1}{2} \ket{3}{4}}.
	\ee	
	Using the conventions for the colour-ordered Feynman rules from \cite{Dixon:1996wi,Elvang:2013cua,Henn:2014yza} one can transform the fundamental generators $T^a$ of the single and double trace colour structure into adjoint generators $f^{abc}$. It turns out that the entire colour structure of \eqref{eq:YMampl} is encoded precisely in the chain of structure constants 
	\be
			\label{eq:ColourStructure}
		c^{a_1 a_2 a_3 a_4},\quad c^{a_1 a_2 a_4 a_3}\quad \text{and} \quad c^{a_1 a_4 a_2 a_3},
	\ee
which are given by Feynman graphs with box topology, i.e. $c^{a_1 a_2 a_3 a_4} := f^{a' a_1 b'} f^{b' a_2 c'} f^{c' a_3 d'} f^{d' a_4 a'}$. This enables us to write the amplitude \eqref{eq:YMampl} in the form \eqref{eq:GTamplitude} with the YM numerators $n_{1^+ 2^+ 3^+ 4^+}^{\text{YM}}$. The exact relation between both numerators reads\footnote{
	It is also possible to derive the numerator \eqref{eq:YMbox} directly in the structure constant colour basis from the integrand of the maximally helicity violating (MHV) $\mathcal{N}=4$ super YM amplitude. The authors of \cite{Bern:1993qk} conjectured a formula which relates at one-loop all-plus primitive amplitudes  of YM to MHV primitive amplitudes of $\mathcal{N}=4$ super YM. This formula also generalizes to higher multiplicity. We thank Radu Roiban for pointing this out to us.}
	\begin{eqnarray}
			\label{eq:YMbox}
		n_{1^+ 2^+ 3^+ 4^+}^{\text{YM}} = 4 a_{1^+ 2^+ 3^+ 4^+}^{\text{YM}}.
	\end{eqnarray}		
	Furthermore the numerators obey the following symmetry properties
	\be
			\nonumber
		n_{i^+ j^+ k^+ l^+}^{\text{YM}} & = & n_{i^+ j^+ l^+ k^+}^{\text{YM}},
			\\
			\label{eq:nYMSym}
		n_{i^+ j^+ k^+ l^+}^{\text{YM}} & = & n_{k^+ l^+ i^+ j^+}^{\text{YM}},
			\\
			\nonumber
		n_{i^+ j^+ k^+ l^+}^{\text{YM}} & = & n_{j^+ k^+ i^+ l^+}^{\text{YM}} = n_{k^+ i^+ j^+ l^+}^{\text{YM}}.
	\ee	
	These properties imply that for all 24 positions of the external legs the numerator reads as \eqref{eq:YMbox}. Once the symmetries of the numerator are known it is trivial to prove that they obey colour-kinematic duality (CKD). CKD demands the relation \eqref{eq:CKD} to hold which in this case translates into
	\begin{eqnarray}
			\label{jacobc}
		c^{a_1 a_2 a_3 a_4} - c^{a_1 a_2 a_4 a_3}  & = & c^{\text{triangle}} :=  f^{a' a_1 b'} f^{b' a_2 c'} f^{c' a' d'} f^{d' a_4 a_3}  
			\\
			\label{eq:BoxYM}
	\implies	n_{1^+ 2^+ 3^+ 4^+}^{\text{YM}} - n_{1^+ 2^+ 4^+ 3^+}^{\text{YM}} & = & n_{\text{triangle}}^{\text{YM}}.
	\end{eqnarray}
	It has been discussed that $n_{\text{triangle}}^{\text{YM}} = 0$. Besides, we have shown that all numerators with box topology are the same $n_{i^+ j^+ k^+ l^+}^{\text{YM}} = n_{1^+ 2^+ 3^+ 4^+}^{\text{YM}}$. These facts imply that \eqref{eq:BoxYM} is trivially obeyed.
	An alternative way to obtain BCJ numerators using light-cone Feynman rules for the
	one-loop rational pure YM amplitudes was introduced in \cite{Boels:2013bi}.

	Thus, in restricting ourselves to the all-plus sector, only the colour structures \eqref{eq:ColourStructure} appear for the YM amplitudes and their numerators satisfy CKD automatically. Therefore to obtain integrands of \theory\,  we have to collect all diagrams of \YMS\, theory which have the same colour structures as \eqref{eq:ColourStructure} and multiply the two numerators divided by the stripped propagators. More details of the DC construction are given in \ref{sec:LagrangianFMRules}.
	
After we have analyzed the first gauge theory let us turn to the second one, i.e. \YMS. The corresponding amplitude of \YMS\, at one-loop reads
	\begin{eqnarray}
			\label{eq:YMScaAmp}
		\left. \mathcal{A}^{1-\text{loop}}_4 \right|_{g^4 \lambda^n} = g^{4} \lambda^n \sum_{\mathcal{S}_m} \sum_{j\in \Gamma}
		\int \frac{d^{4} l}{ (2\pi)^{4} } \frac{1}{S_j} \frac{c_j n_j^{\text{\YMS}}}{\prod_{\alpha_j} D_{\alpha_j}}.
	\end{eqnarray}
It follows from the Lagrangian \eqref{eq:YMSCLagrangaian} that the $\phi^3$ interaction in \YMS\, is proportional to the coupling constants $g \lambda$. Thus the exponent $n \leq 4$ indicates how often the $\phi^3$-interaction appears in the Feynman diagrammatic decomposition of the amplitude. $ n_j^{\text{\YMS}}$ shall be computed by Feynman diagrams generated from the Feynman rules of the theory given in appendix \ref{sec:LagrangianFMRules}.

Once we know both gauge theory numerators we can determine the gravity amplitude via
	\begin{eqnarray}
			\label{eq:YMDCGravity}
		\left. \mathcal{M}^{1-\text{loop}}_4 \right|_{\kappa^{4-n} \gym^n} & = &  \left(\frac{\kappa}{4}\right)^{4-n} \gym^n \sum_{\mathcal{S}_m}
		\int \frac{d^{4} l}{ (2\pi)^{4} } n^{\text{YM}} \sum_{j\in \Gamma_g} \frac{1}{S_j} \frac{n_j^{\text{\YMS}}}{\prod_{\alpha_j} D_{\alpha_j}}.
	\end{eqnarray}
In \cite{Chiodaroli:2014xia,Chiodaroli:2017ngp} it has been deduced how the global colour structure constant $F^{ABC}$ of \YMS\, maps into the local one $f^{abc}$ of \theory. Furthermore, it follows from \eqref{eq:YMScaAmp} and \eqref{eq:YMDCGravity} that the mapping of the coupling constants from the gauge theories to \theory\, is given by %
	\be
			\label{eq:mappconv}
		F^{ABC}  \rightarrow  f^{abc}, 
			\qquad
		(g^2,\lambda)  \rightarrow \left(\frac{\kappa}{4}, 4 \frac{\gym}{\kappa} \right).
	\ee

Thus the first step to obtain the full result in four dimensions for the amplitudes presented in section \ref{sec:Amplitudes} is to evaluate the contributing numerators $n_j^{\text{\YMS}}$ by Feynman diagrams. After the gravity numerators of the amplitude are determined we shall decompose the tensor integral into scalar integrals by using the technique of Veltman-Passarino reduction \cite{Passarino:1978jh}. This feature is nicely implemented in the \texttt{Mathematica} package \texttt{FeynCalc} \cite{Mertig:1990an, Shtabovenko:2016sxi}, which is used in the following calculations. The final scalar Veltman-Passarino functions still depend on the fictitious mass squared $\mu^2$. The next step is then to evaluate the remaining integral over $\mu^2$. However, as the authors of \cite{Bern:1995db} have shown, we can express the integrals containing $\mu^{2}$ as higher dimensional loop integrals, which gives an easy analytic way to determine the four-dimensional amplitude. The relevant formulae to this reduction are given in appendix \ref{sec:regularization}.

\section{Amplitudes}
	\label{sec:Amplitudes}

In this section, all positive helicity amplitudes in \theory\, shall be calculated using the DC construction discussed in section \ref{sec:Procedure}. New results are obtained for 
$\langle 1^+ 2^+ 3^+ 4^{++}\rangle$ and $\langle 1^+ 2^+ 3^{++} 4^{++}\rangle$ at order $\kappa^3$ and $\kappa^4$, respectively. Furthermore, we also present the $\kappa$-corrections at order $\kappa^2$ and $\kappa^4$ for the four-gluon amplitude $\langle 1^+ 2^+ 3^+ 4^{+}\rangle$. Note that a DC expression of these integrands for arbitrary helicity configurations has been given in \cite{Chiodaroli:2014xia} by using the results of \cite{Bern:2013yya}, however, the integrated amplitude has not been published. In the last part of this section we present the result for $\langle 1^{++} 2^{++} 3^{++} 4^{++}\rangle$.
The relevant integrands of \YMS\, which shall be used in the DC prescription are collected in appendix \ref{sec:Integrand}.

\subsection{Amplitudes: $\langle 1^+2^+3^+4^{+}\rangle$}

\begin{figure}
			\center
		\begin{tabular}{c c c}
			\includegraphics[scale=0.7]{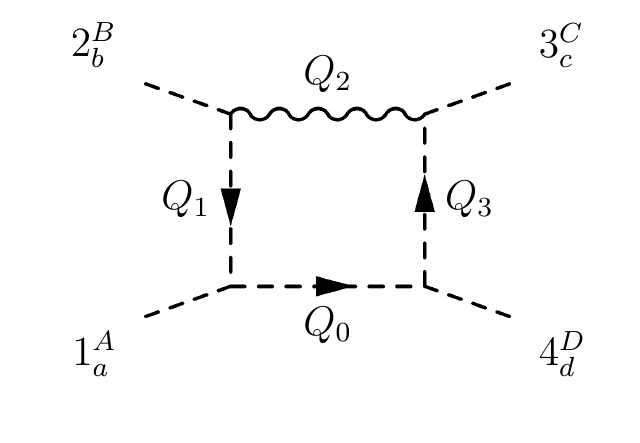}
	&
			\includegraphics[scale=0.7]{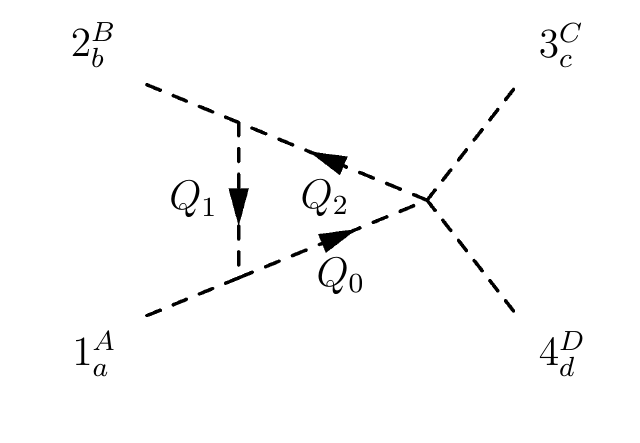}
	&
		\end{tabular}
		\caption{Graphs of these topologies are the only ones that have to be considered in \YMS\, at order $\lambda^2 g^4$ . The other graphs do not contain the colour structures \eqref{eq:ColourStructure}. Curly lines represent propagating gluons and dashed lines represent scalar fields. The internal momenta $Q_i$ are $d$-dimensional.}\label{fig:4gluonkappa2}
	\end{figure}

We shall start the evaluation by calculating gravitional corrections to the four-gluon amplitude. At first we compute the $\kappa^2$ correction to  $\left.\langle 1^+2^+3^+4^{+}\rangle\right|_{\kappa^2}$ with the DC method introduced in the previous section. Therefore we have to calculate in \YMS\ all Feynman graphs which are proportional to $\lambda^2$ according to the DC dictionary presented in \eqref{eq:mappconv}. In section \ref{sec:strategy} we also  pointed out that only the box diagrams are non-vanishing for the all-plus YM amplitude. Hence, only the diagrams which carry the same colour structure as the box diagrams have to be determined in \YMS. A careful analysis shows that only the graph topologies shown in figure \ref{fig:4gluonkappa2} contribute. The integrands in figure \ref{fig:4gluonkappa2} have a very simple form and can be obtained by the Feynman rules given in appendix \ref{sec:LagrangianFMRules}. For example the integrand of the first graph in figure \ref{fig:4gluonkappa2} reads:
	\bee
		\begin{minipage}{4cm}
			\includegraphics[scale=0.6]{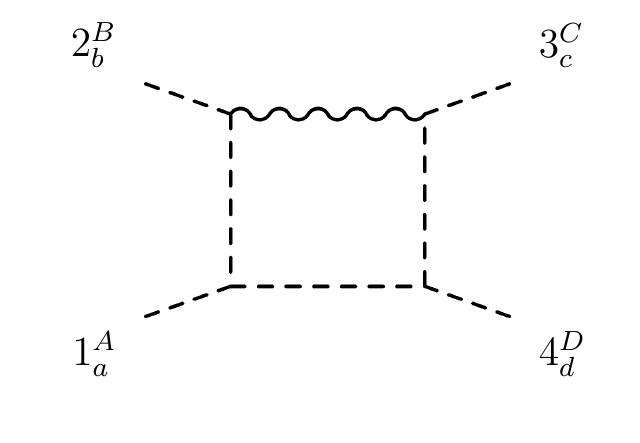}
		\end{minipage} 
		=  \frac{g^4 \lambda^2 c^{abcd}}{D_0 D_1 D_2 D_3} F^{ABE'}F^{E'CD} \left[ \left(p_3 + q_3 \right) \cdot \left( q_1 - p_2 \right) - \mu^2 \right].
	\eee
The factors in the denominator $D_i = Q_i^2 + i \epsilon = q_i^2 -\mu^2 + i \epsilon$ are the denominators of the Feynman propagators. These are defined in appendix \ref{sec:regularization} eq.~\eqref{eq:PropNot}. The colour structure reads $c^{abcd}=  f^{a'ab'}  f^{b'bc'} f^{c'cd'} f^{d'da'}$ and the global SU$(N)$ group information is encoded in $F^{ABE'}F^{E'CD}$, which will be mapped into the adjoint gauge group generators of \theory\ in the
DC.

The amplitude representation \eqref{eq:YMScaAmp} only contains cubic graphs, however, we see that in figure \ref{fig:4gluonkappa2} Feynman graphs with quartic vertices also appear, e.g. the first graph in the second line of figure \ref{fig:4gluonkappa2} reads
	\bee
		\begin{minipage}{4cm}
			\includegraphics[scale=0.6]{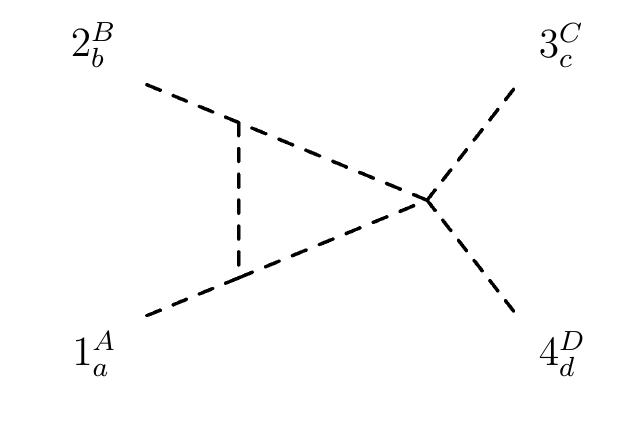}
		\end{minipage} 
		& = & \frac{g^4 \lambda^2}{D_0 D_1 D_2} 
		\left[
			c^{abcd} \left( F^{ACE'}F^{E'DB} -\delta^{CD} F^{A'AB'}F^{B'BA'} \right)
		\right.
				\\
		& & \left.	
		 	+ c^{abdc} \left( F^{ADE'}F^{E'CB} -\delta^{CD} F^{A'AB'}F^{B'BA'} \right)
			+ c^{\text{triangle}} \ldots 
		\right].	
	\eee
We use the Jacobi-identity $c^{abdc} \equiv c^{abcd} + c^{\text{triangle}}$ of (\ref{jacobc}) in our computations which implies that we have to add the numerators of the first two terms. However, we also know that all numerators in YM theory associated to colour structures different from \eqref{eq:ColourStructure} vanish. This implies that the parts of the integrand which effectively contribute to \theory\, are
	\bee
		\begin{minipage}{4cm}
			\includegraphics[scale=0.6]{pics/graph7.pdf}
		\end{minipage} 
		& \effequal & \frac{g^4 \lambda^2}{D_0 D_1 D_2} 
			c^{abcd} \left( F^{ACE'}F^{E'DB} +  F^{ADE'}F^{E'CB} \right.
				\\
		& & \left.	
		 	  - 2 \delta^{CD} F^{A'AB'}F^{B'BA'} \right).
	\eee
The next step is to insert $\frac{D_3}{D_3}$ which makes the graph trivalent formally such that the gravity integrand can be obtained by \eqref{eq:YMDCGravity}.

After collecting all the non-vanishing contributions and building up the complete integrand, we Veltman-Passarino reduce  the amplitude with the \texttt{Mathematica} package \texttt{FeynCalc} \cite{Mertig:1990an, Shtabovenko:2016sxi}. Using the mapping conventions given in \eqref{eq:mappconv} the full amplitude at order $\kappa^2$ reads
		\begingroup
		\allowdisplaybreaks[0]
	\be
			\nonumber
		\mathcal{M}\left.(1^+2^+3^+4^+)\right|_{\kappa^2 \gym^2} & = & 
		-\frac{i}{\left(4 \pi \right)^2} \left(\frac{\kappa \gym}{4}\right)^2 \frac{4}{3}
		\frac{\bra{1}{2} \bra{3}{4}}{\ket{1}{2} \ket{3}{4}} 
			\left(
				 4 f^{abe'}f^{e'cd} \left(U-T \right) + 4 f^{ade'}f^{e'bc} \left(S-U \right) 
			\right.
				\\
				\nonumber
			& & \left.
				+ 4 f^{ace'}f^{e'db} \left(T-S \right) +N S \; \delta^{ab} \delta^{cd} +N T \; \delta^{ad} \delta^{cb} +N U \; \delta^{ac} \delta^{bd} 
			\right)
				\\
			& = &  - i \frac{\kappa^2 \gym^2 }{192 \pi^2} \frac{\bra{1}{2} \bra{3}{4}}{\ket{1}{2} \ket{3}{4}} 
			\left(  4 f^{abe'}f^{e'cd} \left(U-T \right) +N S \; \delta^{ab} \delta^{cd} + \text{perm} \right),
	\ee
		\endgroup
where perm indicates the permutations of the legs 2 and 3 as well as 2 and 4. The kinematical dependence is encoded in the spinor brackets and the Mandelstam variables $S = \ket{1}{2} \bra{2}{1}$, $T = \ket{2}{3} \bra{3}{2}$ and $U = \ket{1}{3} \bra{3}{1}$.

The next correction term  $\left.\langle 1^+2^+3^+4^{+}\rangle\right|_{\kappa^4}$ can be obtained by the same technique as for $\left.\langle 1^+2^+3^+4^{+}\rangle\right|_{\kappa^2 \gym^2}$. All the graphs which contribute are depicted in figure \ref{fig:4gluonkappa4}. The numerators are fairly simple and are listed in the appendix \ref{sec:Integrand4scalar}. A straight forward calculation gives the following integrated amplitude
	\be
			\nonumber
		\mathcal{M}\left.(1^+_a, 2^+_b,3^{+}_c,4^{+}_d)\right|_{\kappa^4} 
		& = & \frac{i}{\left(4 \pi \right)^2} \frac{\kappa^4}{4^4}
		\frac{4}{15}\frac{\bra{1}{2} \bra{3}{4}}{\ket{1}{2} \ket{3}{4}} 
		\left(
			\delta^{ab} \delta^{cd} \left(  40 \; T U -\left( 2 + N_g\right) S^2 \right)
		\right.
			\\
			\nonumber
		&& \left.
			+ \delta^{ac} \delta^{bd} \left( 40 \; S T - \left( 2 + N_g\right) U^2 \right)
			+ \delta^{ad} \delta^{bc} \left( 40 \; S U -\left( 2 + N_g\right) T^2  \right)
		\right)
			\\
		& = & 
		\frac{i }{\left(16 \pi \right)^2} \frac{\kappa^4}{60}
		\frac{\bra{1}{2} \bra{3}{4}}{\ket{1}{2} \ket{3}{4}} 
		\left[
			\delta^{ab} \delta^{cd} \left( 40 \; T U - \left( 2 + N_g\right) S^2  \right) + \text{perm}
		\right].
	\ee
	where again perm indicates the permutations of legs 2 and 3 as well as 2 and 4 and $N_g=\delta^{a'a'}=N^2-1$ is the number of adjoint generators of the Lie algebra.
	\begin{figure}
			\center
		\begin{tabular}{c c c c}
			\includegraphics[scale=0.7]{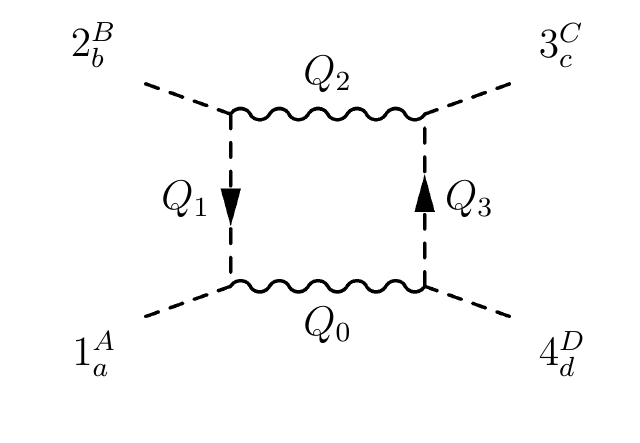}
	&
			\includegraphics[scale=0.7]{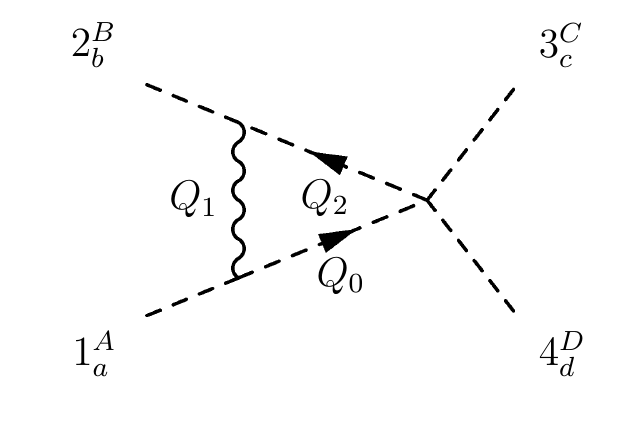}
	&
			\includegraphics[scale=0.7]{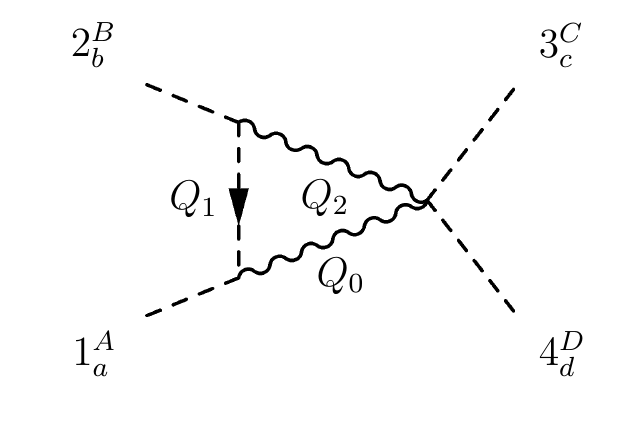}
	\\
			\includegraphics[scale=0.7]{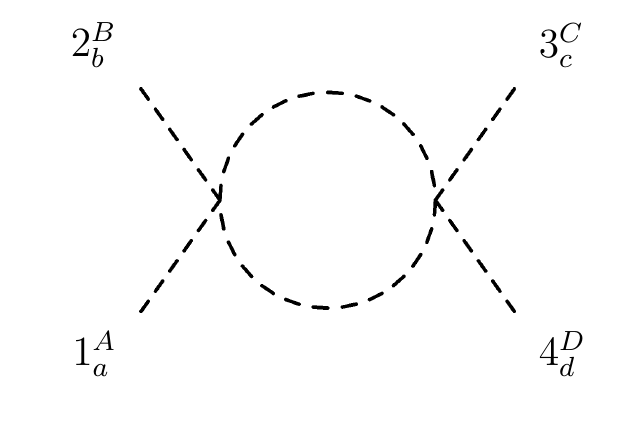}
	&
			\includegraphics[scale=0.7]{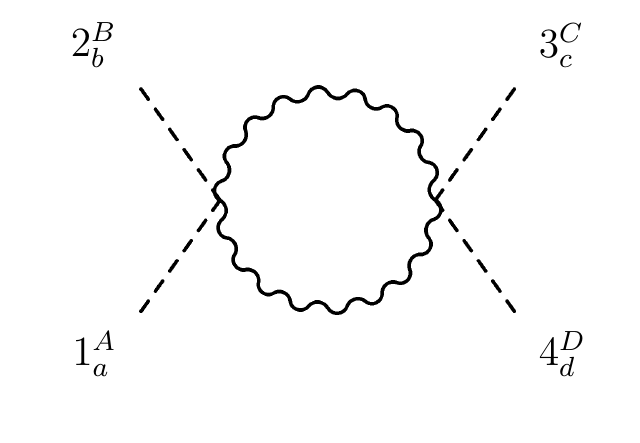}
		\end{tabular}
		\caption{At order $g^4$ the box, triangle and bubble topologies are non-vanishing after double copying.}\label{fig:4gluonkappa4}
	\end{figure}

\subsection{Amplitudes: $\langle 1^+2^+3^+4^{++}\rangle$}

It has been explicitly shown that $\left.\langle 1^+2^+3^+4^{++}\rangle\right|_{\kappa \gym^3}$ vanishes in four dimensions \cite{Nandan:2018ody} using both generalized unitarity and the DC. Therefore the DC calculation is not reproduced in this paper.

\begin{figure}
			\center
		\begin{tabular}{c c c c}
			\includegraphics[scale=0.55]{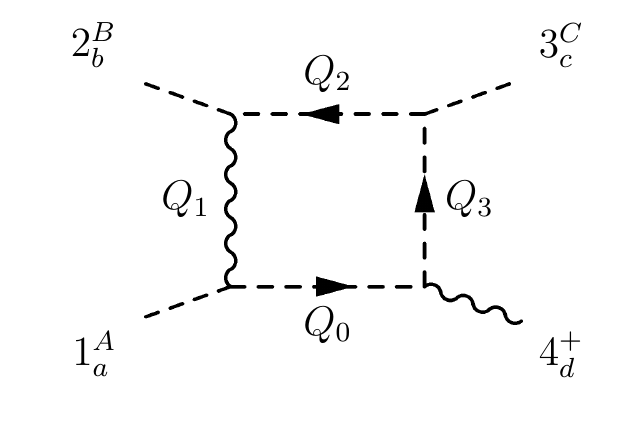}
	&
			\includegraphics[scale=0.55]{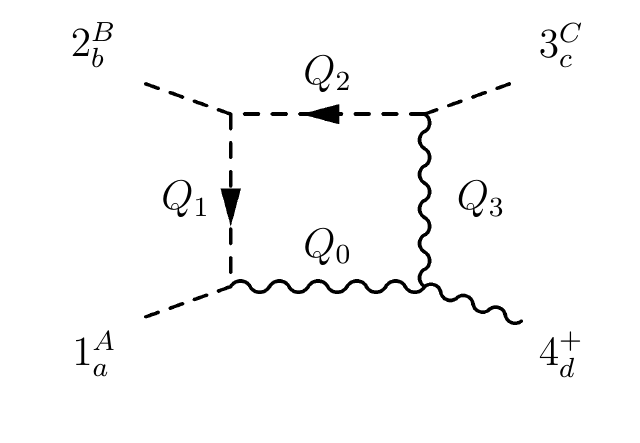}
	&
			\includegraphics[scale=0.55]{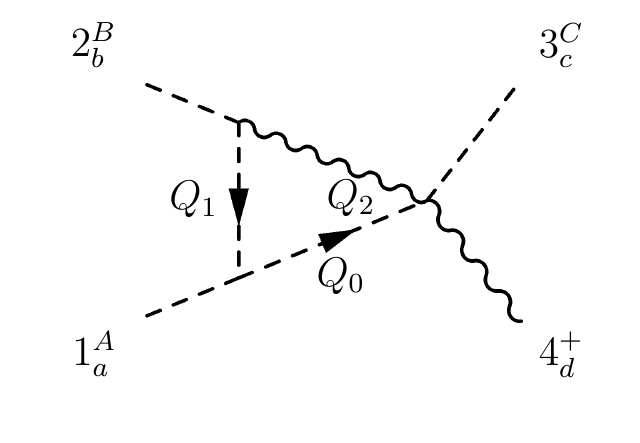}
	&
			\includegraphics[scale=0.55]{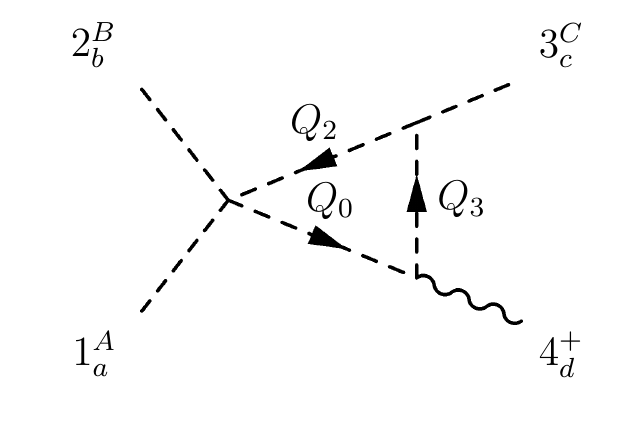}
		\end{tabular}
		\caption{To obtain $\left.\langle 1^+2^+3^+4^{++}\rangle\right|_{\kappa^3 \gym}$ we have to analyze this type of graphs in \YMS.}\label{fig:3gluonkappa3}
	\end{figure}

We therefore move on to the order $\kappa^{3}$ contribution to this amplitude. It can be proven that for the last graph in figure \ref{fig:3gluonkappa3} all box colour structures vanish after applying Jacobi identites.

A typical integrand of figure \ref{fig:3gluonkappa3} is of the form
	\bee
		\begin{minipage}{4cm}
			\includegraphics[scale=0.6]{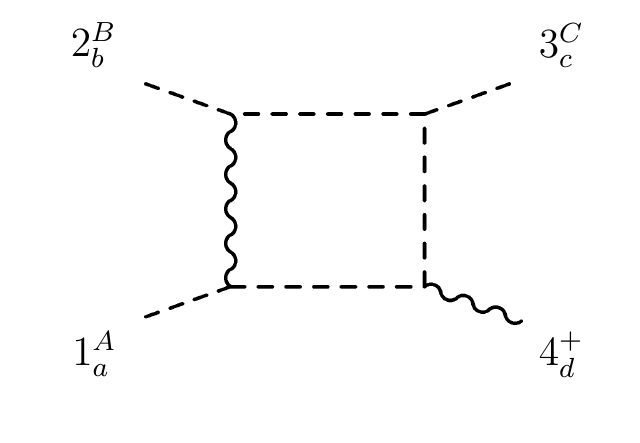}
		\end{minipage} 
		& = & \frac{-i g^4 \lambda}{D_0 D_1 D_2 D_3} c^{abcd} F^{ABC}
		\left[  \left(p_2 +q_2\right) \cdot \left(q_0 -p_1 \right) -\mu^2 \right]
		\frac{\ketbra{r_4}{q_0+q_3}{4}}{\sqrt{2} \ket{r_4}{4}},
	\eee
where four-dimensional spinor helicity variables are used to represent the polarization vector. Compared to the previous two amplitudes we have now a gauge choice that is encoded in the reference vector $r_4$ which can be chosen arbitrarily but not such that it is proportional to $p_4$. We have done the calculation with the choices $r_4 \in \{p_1,p_2,p_3\}$. Since the amplitude has to be invariant under different gauges, this represents a powerful crosscheck for the final result.

After all integrands of \YMS\, have been determined one can construct the gravity integrand using \eqref{eq:YMDCGravity}. Evaluating and reducing this expression yields the simple result
	\begin{eqnarray}
		\left. \mathcal{M}(1^+_a, 2^+_b,3^{+}_c,4^{++}) \right|_{\gym \kappa^3} =  - \frac{\kappa^3 \gym}{(8\pi)^2}  \frac{f^{abc}}{\sqrt{2}} \frac{\bra{4}{1} \bra{4}{2} \bra{4}{3} \bra{1}{2}}{\ket{3}{4}}.
	\end{eqnarray}		

\subsection{Amplitudes: $\langle 1^+2^+3^{++} 4^{++}\rangle$}

\begin{figure}
			\center
		\begin{tabular}{c c}
			\includegraphics[scale=0.7]{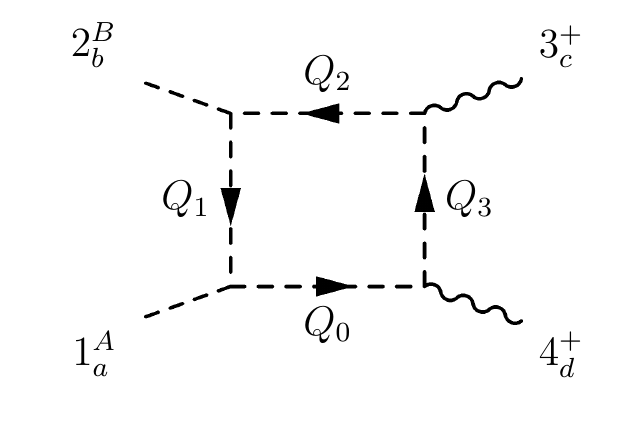}
	&
			\includegraphics[scale=0.7]{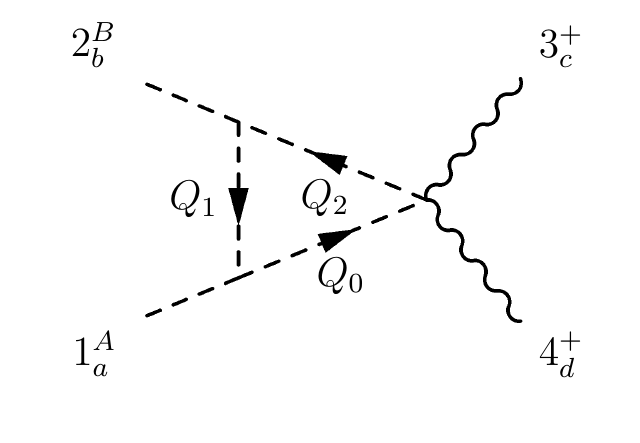}
		\end{tabular}
		\caption{This pair of graph topologies represent graphs which appear at leading order in $\lambda^2 g^2$ for $\langle 1^+2^+3^{++} 4^{++}\rangle$.}\label{fig:2gluonkappa2}
	\end{figure}

The leading order in $\kappa^2$ for the  amplitude $\left.\langle 1^+2^+3^{++} 4^{++}\rangle\right|_{\kappa^2 \gym^2}$ has been determined in \cite{Nandan:2018ody} using the unitarity based two cut method. We shall see that with the DC prescription \eqref{eq:YMDCGravity}  this result is much easier obtained. Only the graph topologies drawn in figure \ref{fig:2gluonkappa2} have to be evaluated on the \YMS\, side. Besides, the calculation can even further be simplified by choosing the reference momenta $r_i$ for the gluon polarization vectors at the legs three and four to be the same such that the integrands containing the quartic vertex give zero identically.%
	\footnote{This follows from $\epsilon_3^+ \cdot \epsilon_4^+ \sim \ket{r_3}{r_4}$.}

Thus only the first type of graph from figure \ref{fig:2gluonkappa2} has to be determined. The resulting amplitude is given by
	\begin{eqnarray}
		\left. \mathcal{M}(1^+_a, 2^+_b,3^{++},4^{++}) \right|_{\gym \kappa^3} =\frac{i}{(4\pi)^2} \left(\frac{\kappa \gym}{2}\right)^2 f^{a'a b'} f^{b' b a'} \frac{S}{6} \frac{\bra{1}{2} \bra{3}{4}^2}{\ket{1}{2} \ket{3}{4}^2}.
	\end{eqnarray}		

This result agrees with the expression given in \cite{Nandan:2018ody} if one inserts the pair of coupling constants $\left(\gym \kappa/2\right)^2$ and the colour structure $N \, \text{Tr}\left(T^a T^b \right) = N \, \delta^{ab}= f^{a'ab'}f^{b'ba'}$ following from the decomposition of a one-loop amplitude into partial amplitudes. Since the result stated in \cite{Nandan:2018ody} is the only partial amplitude which contributes, both expression coincide.

\begin{figure}
			\center
		\begin{tabular}{c c c c}
			\includegraphics[scale=0.55]{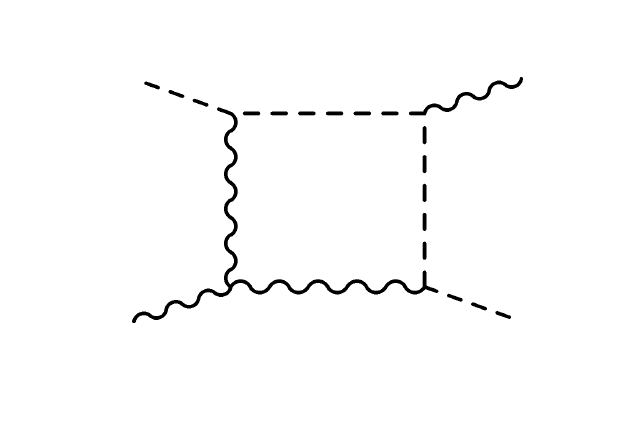}
	&
			\includegraphics[scale=0.55]{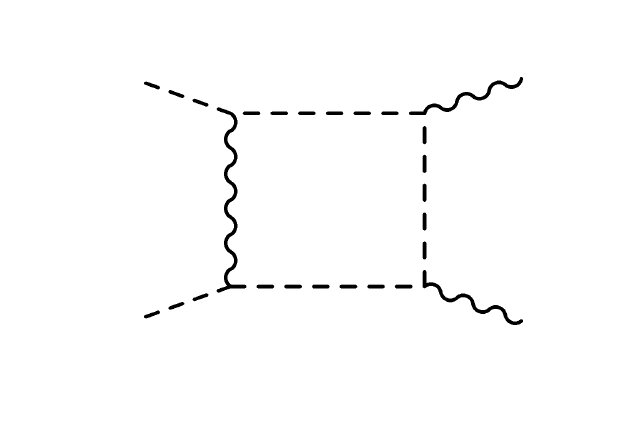}
	&
			\includegraphics[scale=0.55]{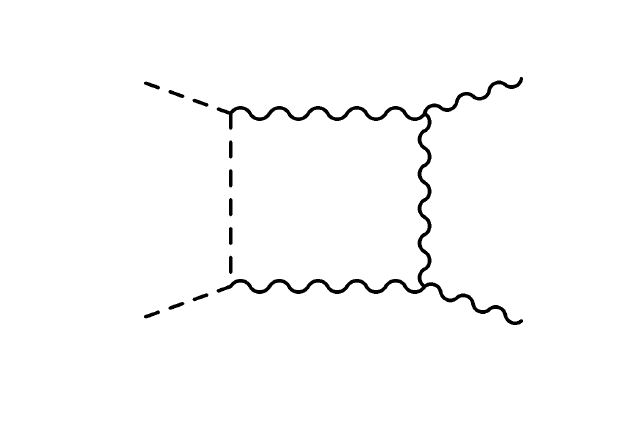}
	&
			\includegraphics[scale=0.55]{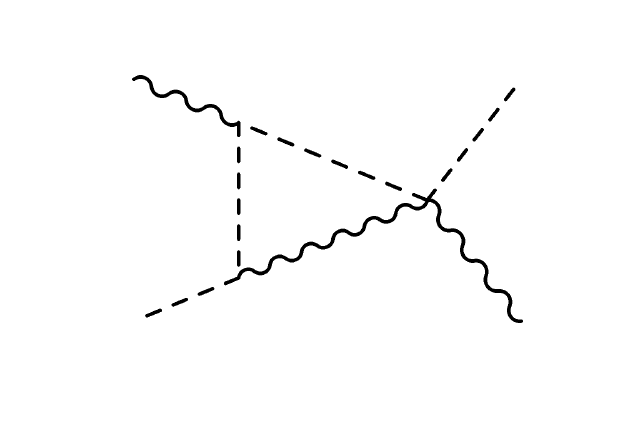}
	\\
			\includegraphics[scale=0.55]{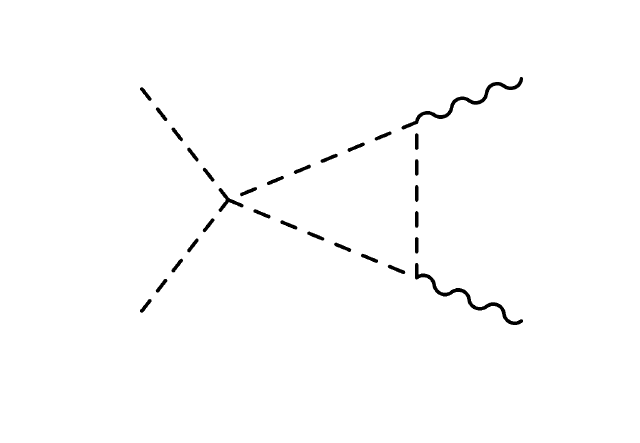}
	&
			\includegraphics[scale=0.55]{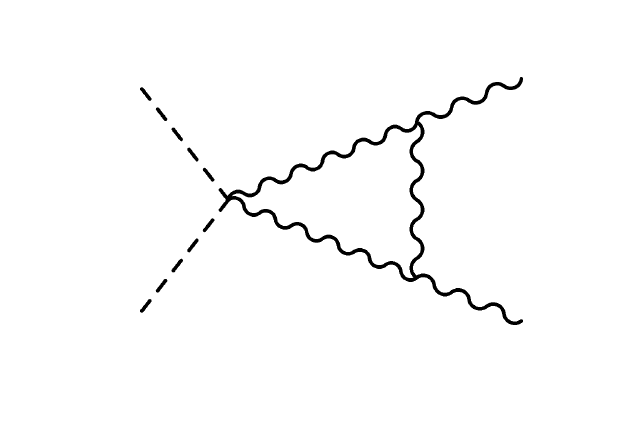}
	&
			\includegraphics[scale=0.55]{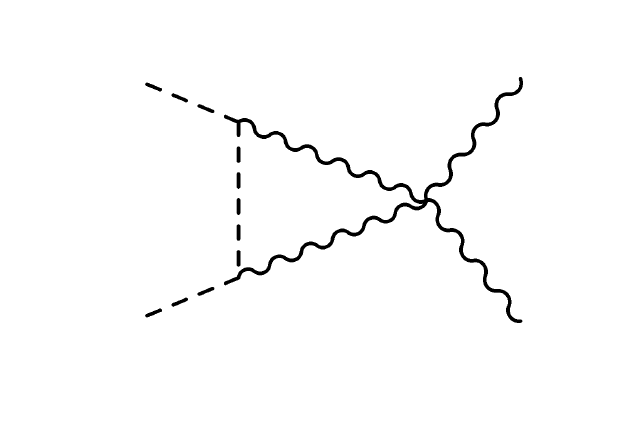}
	&
			\includegraphics[scale=0.55]{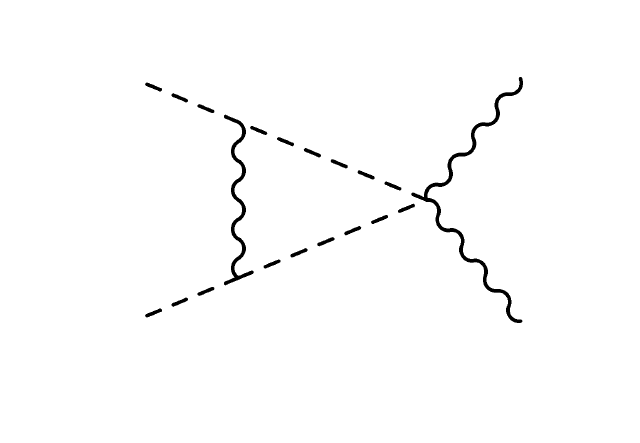}
	\\
			\includegraphics[scale=0.55]{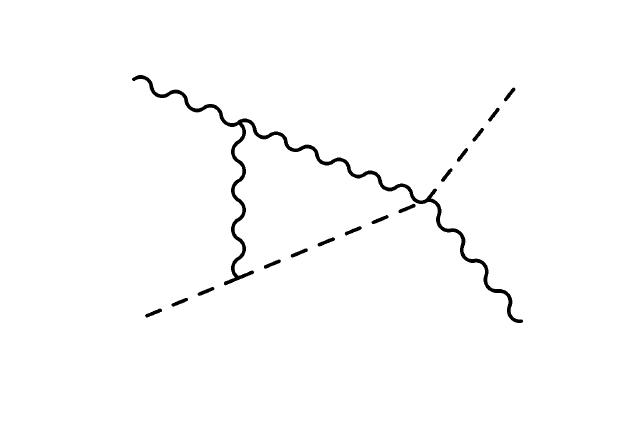}
	&
			\includegraphics[scale=0.55]{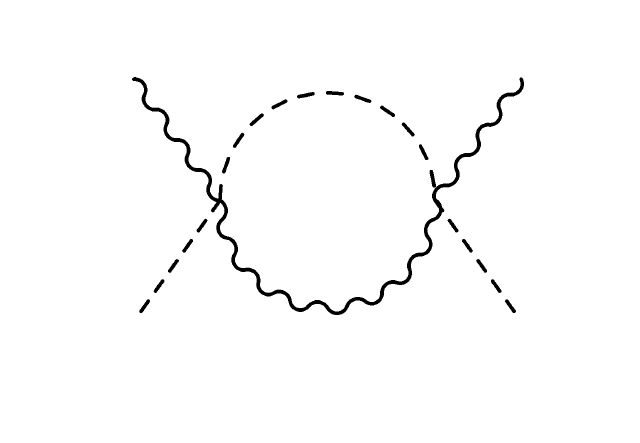}
	&
			\includegraphics[scale=0.55]{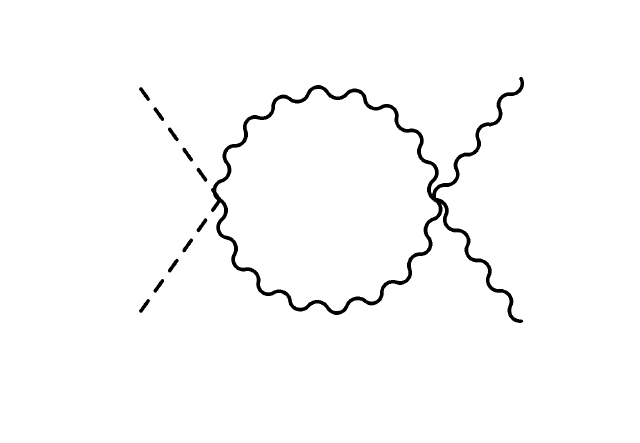}
	&
			\includegraphics[scale=0.55]{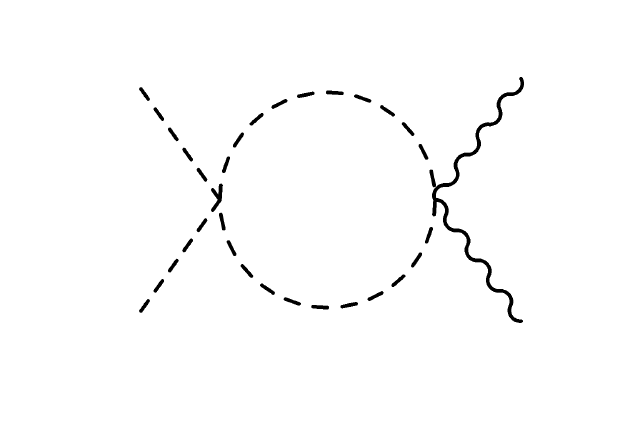}

		\end{tabular}
		\caption{At $g^4$ these graph topologies shall contribute to the gravity amplitude $\left.\langle 1^+2^+3^{++} 4^{++}\rangle\right|_{\kappa^4}$.}\label{fig:2gluonkappa4}
	\end{figure}

The $\kappa^4$ contribution can be determined by the graphs given in figure \ref{fig:2gluonkappa4}. Applying the same steps as before we arrive at the result
	\begin{eqnarray}
		\left. \mathcal{M}(1^+_a, 2^+_b,3^{++},4^{++}) \right|_{\kappa^4} = i \frac{\kappa^4}{\left(16 \pi\right)^2}  \frac{\bra{2}{1}^2 \bra{4}{3}^3}{\ket{3}{4}} \frac{2+N_g}{90}	\delta^{ab}.
	\end{eqnarray}	
Here $N_g$ represents again the dimension of the adjoint representation of the gauge group. This result has been calculated for the following choices of reference momenta $r_3 \in \{p_1,p_2,p_4\}$, ${r_4 \in \{p_1,p_2,p_3\}}$ yielding identical results. 

\subsection{Amplitude: $\langle 1^{++} 2^{++} 3^{++} 4^{++}\rangle$}

\begin{figure}
			\center
		\begin{tabular}{c c c c}
			\includegraphics[scale=0.55]{pics/graph38.pdf}
	&
			\includegraphics[scale=0.55]{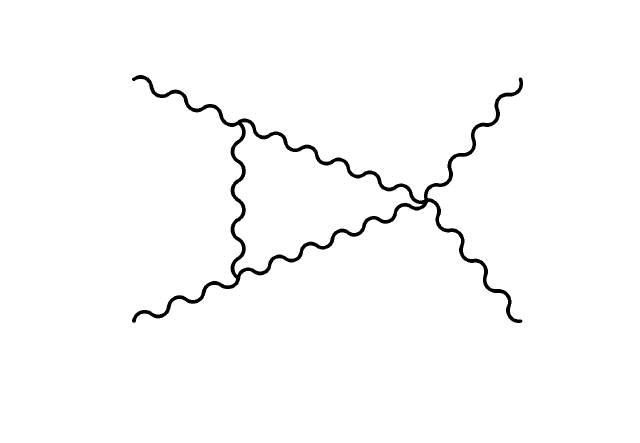}
	&
			\includegraphics[scale=0.55]{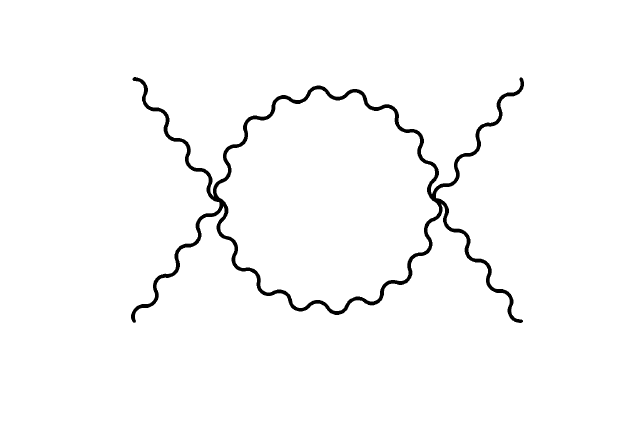}
	\\
			\includegraphics[scale=0.55]{pics/graph41.pdf}
	&
			\includegraphics[scale=0.55]{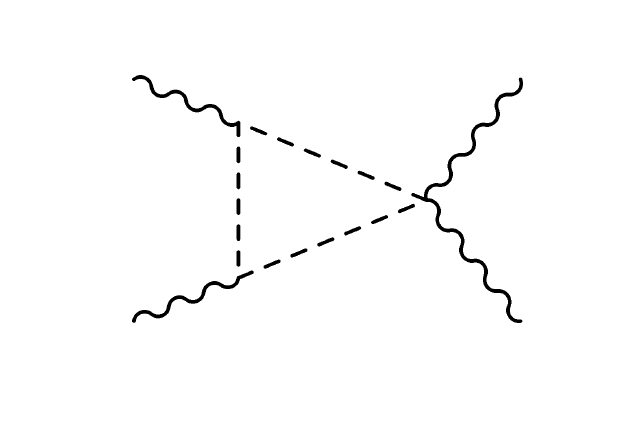}
	&
			\includegraphics[scale=0.55]{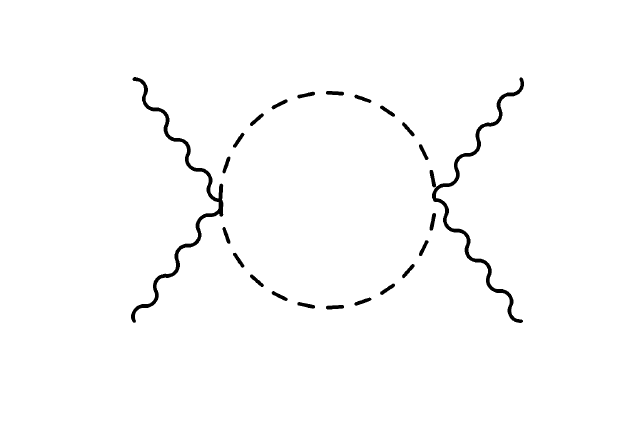}
	&
		\end{tabular}
		\caption{These topologies in \YMS\, have to be evaluated to obtain $\left.\langle 1^{++} 2^{++} 3^{++} 4^{++}\rangle\right|_{\kappa^4}$.\label{fig:0gluonkappa4}}
	\end{figure}

The remaining all-plus amplitude of \theory\, only contains gravitons as asymptotic states. In principle all the diagrams given in figure \ref{fig:0gluonkappa4} contribute to the integrand. It can be seen that the graphs in the first row arise from pure YM theory. It has been discussed in section \ref{sec:strategy} that the calculation of these diagrams simplifies by choosing an appropriate gauge such that the graphs with box topology are the only non-vanishing ones, i.e.~the only integrand which survives is exactly given by \eqref{eq:YMbox}. For example this has been demonstrated in \cite{Brandhuber:2005jw} using unitarity cuts. To simplify their analysis, the propagating gluon has been replaced by a complex scalar which is possible due to the supersymmetric Ward–Takahashi identities. Since a complex scalar has two degrees of freedom we can represent this relation diagrammatically by
	\bee
		\begin{minipage}{5cm}
			\includegraphics[scale=0.7]{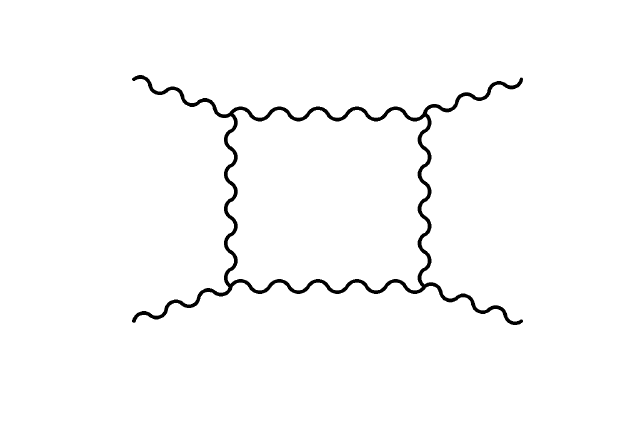}
		\end{minipage} 
		=  \hspace{0.7cm}
		\quad 2 \hspace{-0.7cm}
		\begin{minipage}{5cm}
			\includegraphics[scale=0.7]{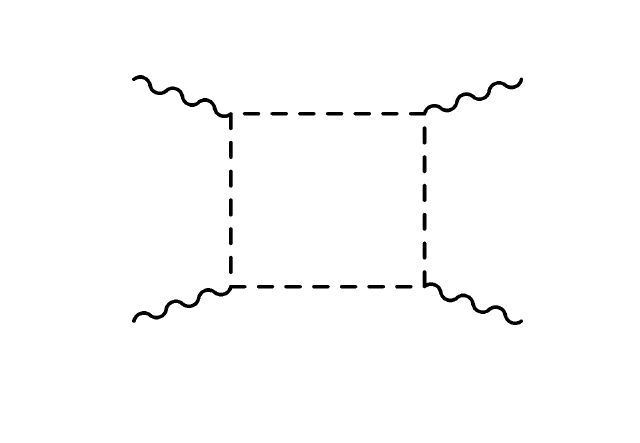}
		\end{minipage}
	\eee
This immediately shows that the graphs in the first and second row are intimately related. Hence we can write the amplitude as the sum of three scalar boxes in $4-2\epsilon$ dimensions, which are defined in equation \eqref{eq:DefInt} of the appendix \ref{sec:regularization}:
		\begingroup
		\allowdisplaybreaks[0]
	\bee
			 \mathcal{M}(1^{++}, 2^{++},3^{++},4^{++})  & = & 
			 \frac{i}{\left(4\pi\right)^{2-\epsilon}} \frac{\kappa^4}{4} 
			\left(\frac{\bra{1}{2} \bra{3}{4}}{\ket{1}{2} \ket{3}{4}}\right)^2			
					\\
			 & & \times \Bigl( I_4\left[\mu^8;S,T\right] + I_4\left[\mu^8;T,U\right] + I_4\left[\mu^8;U,S\right] \Bigr)  \left(1 + \frac{N_g}{2} \right).
	\eee
		\endgroup
In four dimensions the result simply reduces to
	\be
			\mathcal{M}(1^{++}, 2^{++},3^{++},4^{++}) = -\frac{i}{\left(4\pi\right)^{2}} \kappa^4
			\left(\frac{\bra{1}{2} \bra{3}{4}}{\ket{1}{2} \ket{3}{4}}\right)^2
			\frac{S^2 +T^2 + U^2}{1920}
			\left(2 + N_g \right).
	\ee
According to \cite{Johansson:2014zca} we can remove the dilaton and axion by subtracting twice the contribution generated by the adjoint scalar circulating in the loop. For the all-plus amplitude the scalar part is given by
	\bee
			\mathcal{M}^{\text{scalar}}(1^{++}, 2^{++},3^{++},4^{++})  = -\frac{i}{\left(4\pi\right)^{2}} \kappa^4 
			\left(\frac{\bra{1}{2} \bra{3}{4}}{\ket{1}{2} \ket{3}{4}}\right)^2
			\frac{S^2 +T^2 + U^2}{3840}
	\eee
which implies that the pure EYM result reads
	\be
		\label{eq:4graviton}
			\mathcal{M}^{\text{EYM}}(1^{++}, 2^{++},3^{++},4^{++})  = -\frac{i}{\left(4\pi\right)^{2}} \kappa^4
			\left(\frac{\bra{1}{2} \bra{3}{4}}{\ket{1}{2} \ket{3}{4}}\right)^2
			\frac{S^2 +T^2 + U^2}{1920}
			\left(1 + N_g \right).
	\ee
We note that the pure gravity part of \eqref{eq:4graviton} also agrees with \cite{Bern:1998sv, Dunbar:1994bn}.%
	\footnote{Note that $\frac{\bra{1}{2} \bra{3}{4}}{\ket{1}{2} \ket{3}{4}} = - \frac{S T}{\ket{1}{2} \ket{2}{3} \ket{3}{4} \ket{4}{1}}$.}

\section{Conclusions}

	In this paper we have calculated the one-loop corrections to \theory\, for all four point amplitudes with positive helicity configuration. Using the DC of YM and \YMS\, as well as the fact that diagrams with box topology are the only ones on the YM side that have a non-vanishing contribution, we have evaluated all the diagrams of \YMS\, which contain the colour structure of graphs with box topology. Furthermore, we have checked that our results obtained for $\left.\langle 1^+,2^+,3^{++},4^{++}\rangle \right|_{\kappa^2 \gym^2}$ and the pure gravitational part of $\left.\langle 1^{++},2^{++},3^{++},4^{++}\rangle\right|_{\kappa^4}$ coincide with known results in the literature.
	
	Generally, it would be very interesting to be able to obtain the pure \theory\, results from the amplitudes calculated in this paper. An analysis of possible interaction terms generated by the Lagrangian \eqref{eq:EYMLagrangian} shows that the axion and dilaton cannot contribute at one loop for a four point amplitude if the number of asymptotic gluon states matches the power of the coupling constant $g$. The reason is that in this case only the gauge fields can propagate in the loop, because all fields couple at least quadratically to it. Similarly, we were able to seperate the axion and dilaton contribution to the four graviton amplitude from the gluon and graviton contributions. However, all these cases have in common that only one particle type is circulating in the loop. All the remaining amplitudes contain a mixed type of particles circulating in the loop, which makes a subtraction of the axion and dilaton rather difficult. 
	
	Therefore it would be very interesting to remove the axion and dilaton contribution from all the amplitudes to obtain pure \theory\, amplitudes. It might be possible to subtract these fields by a similar method used in \cite{Johansson:2014zca}.
	
	Moreover, one can compute the remaining amplitudes with arbitrary helicity configurations at four points. These can be obtained by using the DC method in the same way as it has been done here because in \cite{Bern:2013yya} the colour-kinematics representation for the YM amplitudes at one-loop has been presented for arbitrary helicity configurations. However, the calculation is much more tedious since the numerators given in \cite{Bern:2013yya} are non-zero for all possible graph topologies. However, this is left for future work.
\\
\\

\subsection*{Acknowledgments}

We thank Henrik Johansson, Radu Roiban and Gabriele Travaglini for comments and discussions. The work of JP is supported through funds of Humboldt-University Berlin in the framework of the German excellency initiative. JF would like to acknowledge the group members of the groups "Mathematical Physics of Space, Time and Matter" and "Quantum Field and String Theory" for discussions and especially Sourav Sarkar for reading the manuscript, suggestions and discussions. Moreover, we would like to thank the CERN theory department for hospitality where this work was partly completed.

\newpage

	\appendix
\section{Appendix}

\subsection{Regularization}
	\label{sec:regularization}
	To regularize the Feynman integrals we shall use four-dimensional helicity (FDH) regularization. This is a dimensional regularization scheme in which fermions and gauge bosons of observed and unobserved particles have two helicity states and the momenta of the observed particles are also kept in four dimensions. Only the momenta of unobserved particles are continued to $4-2\epsilon$ dimensions with $\epsilon <0$ \cite{Bern:1991aq, Bern:2002zk, Bern:1995db}. These properties make this scheme in particular useful for expressing the amplitudes in spinor helicity formalism. 
		
Thus we separate the $d$-dimensional vector $L := (l_{4},l_{-2\epsilon}):= (l,\mu)$, where the two vector spaces are orthogonal $ l \cdot \mu = 0$. Therefore we have $L^2 = l^2 -\mu^2$ using the four-dimensional mostly minus convention $(+,-,-,-)$. So we can view a higher dimensional vector as a lower dimensional vector whose mass-squared is shifted by $\mu^2$.%
	\footnote{We have $L^2 = l^2 -\mu^2$ because we have mostly minus signature and therefore the $-2\epsilon$ should have only minus signature.
		If $L^2= M^2$ then $l^2 = L^2 + \mu^2 = M^2 + \mu^2$.}

	In particular a scalar product with a four-dimensional vector projects always on the four-dimensional vector space e.g. $\varepsilon_4 \cdot L = \varepsilon_4 \cdot l$. This implies one can treat the loop integration with "massless" loop momentum $L$ in $d$ dimensions as a massive loop momentum $l$ in four dimensions. Since we are dealing with four external particles and the dimension of integration is $d>4$, one can think of all external particles lying in a four-dimensional subspace. Thus the external particles can always be represented by spinor helicity variables.
	
	Following the conventions of \cite{Nandan:2018ody} the $d$-dimensional scalar Feynman integrals are defined by the expression
	\be
			\label{eq:DefInt}
		\frac{i}{\left(4\pi\right)^{2-\epsilon}}I_n\left[\mu^{2r}\right] :=
		\int \frac{d^{4}l}{\left(2\pi\right)^4} \int\frac{d^{-2\epsilon} \mu}{\left(2\pi\right)^{-2\epsilon}} 
		\frac{\mu^{2r}}{ D_0 \cdots D_{n-1}},
	\ee
where 
	\be
			\label{eq:PropNot}
		D_i = Q_i^2 + i \epsilon = q_i^2 - \mu^2 + i \epsilon, \qquad 
		q_j = l + \sum_{i=1}^{j} p_i.
	\ee
Using the formula
	\be
			\label{eq:RemMu}
		I_n^{d=4-2\epsilon} \left[ \mu^{2r} \right] 
		= - \epsilon \left(1-\epsilon\right) \left(2-\epsilon \right) \cdots \left(r-1-\epsilon \right) I_n^{d=4+2r -2\epsilon} \left[ 1 \right],
	\ee
which can be derived by transforming the $d^{-2\epsilon}\mu$-integration into spherical polar coordinates \cite{Bern:1995db}. Hence formula \eqref{eq:RemMu} removes the dependence on the fictitious mass $\mu^{2r}$ parameter by shifting the dimension of the integration variable to $4+2r-2\epsilon$.

From the $4-2\epsilon$-dimensional expression for the bubble, the one-mass triangle
	\bee
		I_2 \left[1;S\right] & = & r_{\Gamma} \frac{\left(-S\right)^{-\epsilon}}{\epsilon \left(1-2\epsilon\right)}, \qquad
		I_3\left[1;S\right]  =  - \frac{r_{\Gamma}}{\epsilon^2} \left( -S \right)^{-1-\epsilon} 		
			\\
		\text{with} \qquad r_{\Gamma} & := & \frac{ \Gamma\left(1+\epsilon \right) \Gamma^2 \left(1 - \epsilon \right)}{\Gamma\left(1-2\epsilon \right)}
	\eee
and the zero-mass box
	\bee
		I_4 \left[1;S, T \right] & = & r_{\Gamma} \frac{2}{S T}
			\left[
				\frac{\left(-S\right)^{-\epsilon}}{\epsilon^2} {}_{2}F_1 \left(1,-\epsilon, 1-\epsilon; 1+\frac{S}{T}\right)
				+ \frac{\left(-T\right)^{-\epsilon}}{\epsilon^2} {}_{2}F_1 \left(1,-\epsilon, 1-\epsilon; 1+\frac{T}{S}\right)
			\right]
	\eee
one can derive exact expressions in arbitrary dimensions. The reason is that  these expression are exact to all orders in $\epsilon$ such that one obtains higher dimensional integrals by simply shifting the value of $\epsilon$:
	\bee
		&& d= 6-2 \epsilon \; \text{is obtained by } \epsilon \rightarrow \epsilon -1,
			\\
		&& d= 8-2 \epsilon \; \text{is obtained by } \epsilon \rightarrow \epsilon - 2,
			\\
		&& \text{etc.},
	\eee
which follows from the expression \eqref{eq:DefInt}. Together with the formula \eqref{eq:RemMu}, all scalar Veltman-Passarino functions can be expressed in terms of Mandelstam variables $S= \ket{1}{2} \bra{2}{1}$, $T=\ket{1}{4} \bra{4}{1}$, and $U=\ket{1}{3} \bra{3}{1}$.
	
	We shall use the following explicit expressions in four dimensions:
	\be
			\nonumber
		I_2 \left[ \mu^2;S\right] & = & - \frac{S}{6} + \mathcal{O}(\epsilon), \qquad \quad
		I_2 \left[\mu^4; S \right] = -\frac{S^2}{60} + \mathcal{O}(\epsilon),
			\nonumber			
			\\
		I_2 \left[\mu^6; S \right] & = & - \frac{S^3}{420} + \mathcal{O}(\epsilon),
			\nonumber
			\\
		I_3\left[\mu^2; S \right] & = & \frac{1}{2} + \mathcal{O}(\epsilon), \qquad \qquad
		I_3\left[\mu^4; S \right] = \frac{S}{24} + \mathcal{O}(\epsilon),
			\nonumber
			\\
		I_3\left[\mu^6; S \right] & = & \frac{S^2}{180} +\mathcal{O}(\epsilon),
			\label{eq:ScalarInt}
			\\
		I_4 \left[\mu^2; S, T \right] & = &  \mathcal{O}(\epsilon), \qquad \qquad \qquad
		I_4 \left[\mu^4; S, T \right]= -\frac{1}{6} + \mathcal{O}(\epsilon),
			\nonumber
			\\
		I_4 \left[\mu^6; S, T \right] & = & - \frac{S+T}{60} + \mathcal{O}(\epsilon), \quad 
		I_4 \left[\mu^8; S, T \right] = - \frac{1}{840} \left(2 S^2 + S T + 2 T^2 \right) + \mathcal{O}(\epsilon).
			\nonumber
	\ee

\subsection{Feynman rules}
	\label{sec:LagrangianFMRules}

We shall give a short review of how we can determine the actual gravity amplitude from the DC of the two theories \eqref{eq:Lagr}. It turns out that in this case the gravity theory is uniquely determined by its spectrum and its cubic coupling, i.e. its three point amplitude. Scrutinizing the spectrum of the gravity theory $(\text{YM + } \phi^3) \otimes_{\text{DC}} \text{YM}$ one can deduce that it contains a graviton, an antisymmetric two-form and the dilaton. However, the latter two can be removed by introducing ghost fields in the double copy of the loop amplitudes. Thus the next step is to normalize the three point vertex in $\mathcal{L}^{\text{\YMS}}$ appropriately to obtain \theory. This has been done in \cite{Chiodaroli:2014xia,Chiodaroli:2017ngp} by explicitly calculating the three point amplitude of $\mathcal{N}=2$ Einstein-Yang-Mills and then truncating this to the bosonic sector.

\begin{table}[H]
		\caption{Feynman rules for \YMS\, derived from the Lagrangian \eqref{eq:YMSCLagrangaian}. In our conventions all momenta are outgoing. For the gluon propagator the Feynman gauge is used.}		\label{tab:FeynRules}
	\centering
		\begin{tabular}{c l}
			\includegraphics[scale=0.6, align=c]{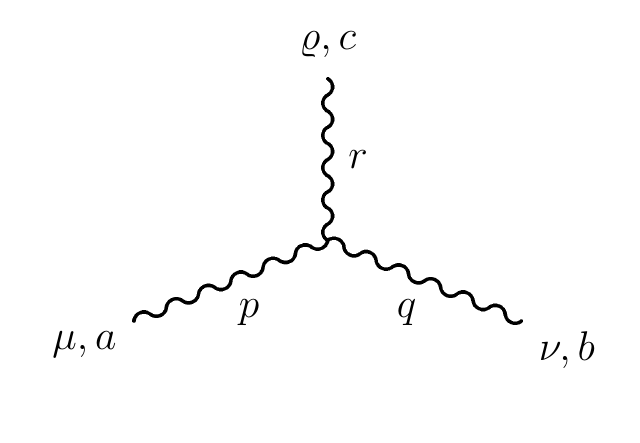} 
				& \hspace{0.55cm}
			$ - g f^{abc} 
			\left[
				\left(r_{\mu} -q_{\mu} \right) \eta_{\nu \varrho} + \left(p_{\nu}-r_{\nu} \right) \eta_{\varrho \mu} 
				+\left(q_{\varrho} -p_{\varrho} \right) \eta_{\mu \nu}
			\right]$
				\\
			\includegraphics[scale=0.6, align=c]{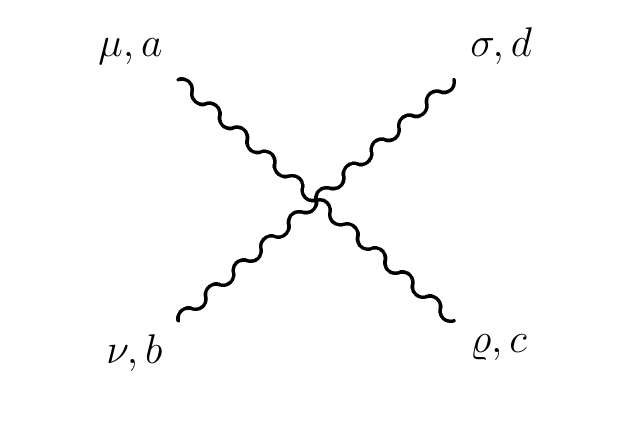}
				&
				\begin{minipage}{0cm}
					\bee
			 &&- i g^2 \left[
				f^{abe'}f^{e'cd} \left( \eta_{\mu \varrho} \eta_{\nu \sigma} - \eta_{\mu \sigma} \eta_{\nu \varrho} \right)
				\right.
				\\
				&&
				+ f^{ace'}f^{e'db} \left( \eta_{\mu \sigma} \eta_{\varrho \nu} - \eta_{\mu \nu} \eta_{\varrho \sigma} \right)
				\\
				&&
				\left.
				+ f^{ade'}f^{e'bc} \left( \eta_{\mu \nu} \eta_{\sigma \varrho} - \eta_{\mu \varrho} \eta_{\sigma \nu} \right)			
			\right]
					\eee
				\end{minipage}	
				\\
			\includegraphics[scale=0.6, align=c]{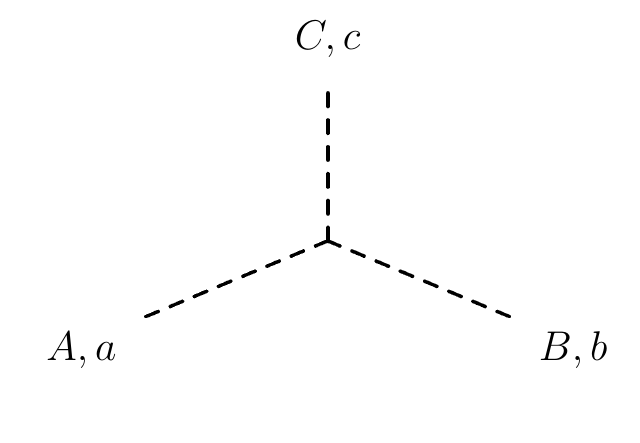}
				& \hspace{0.55cm}
			$i \lambda g f^{abc} F^{ABC}$
				\\
			\includegraphics[scale=0.6, align=c]{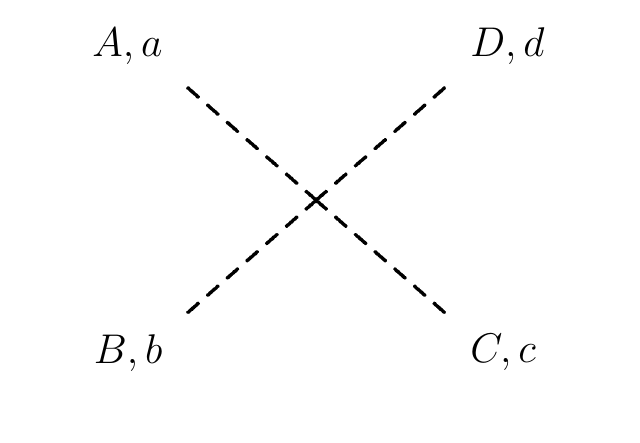}
				&
				\begin{minipage}{0cm}
					\bee
			 &&- i g^2 \left[
				f^{abe'}f^{e'cd} \left( \delta^{AC} \delta^{BD} - \delta^{AD} \delta^{BC} \right)
				\right.
				\\
				&&
				+ f^{ace'}f^{e'db} \left( \delta^{AD} \delta^{BC} - \delta^{AB} \delta^{CD} \right)
				\\
				&&
				\left.
				+ f^{ade'}f^{e'bc} \left( \delta^{AB} \delta^{CD} - \delta^{AC} \delta^{BD} \right)			
			\right]
					\eee
				\end{minipage}	
			
				\\
			\includegraphics[scale=0.6, align=c]{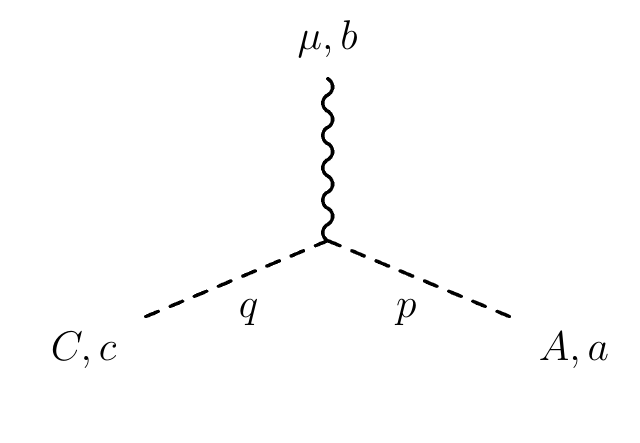}
				& \hspace{0.55cm}
			$ g f^{abc} \delta^{AC} \left( p_{\mu} - q_{\mu} \right)
			$ 	
				\\
			\includegraphics[scale=0.6, align=c]{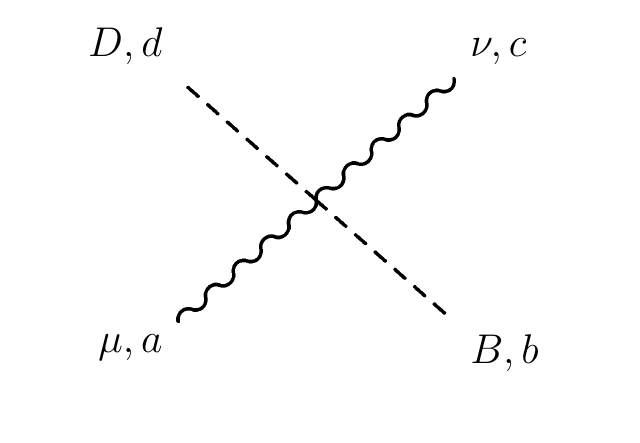}
				& \hspace{0.55cm}
			$  i g^2 \left(f^{abe'}f^{e'cd} + f^{cbe'}f^{e'ad} \right) \delta^{BD} \eta_{\mu \nu}
			$
				\\
			\includegraphics[scale=0.6, align=u]{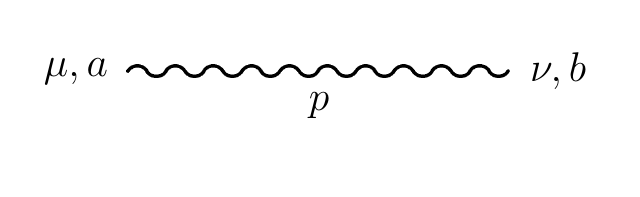}
				& \hspace{0.55cm}
				\begin{minipage}{0cm}	
					\bee
				  		\frac{-i \;\eta^{\mu \nu} \, \delta^{ab}}{p^2+i\epsilon}
					\eee
				\end{minipage}						
				\\
			\includegraphics[scale=0.6, align=u]{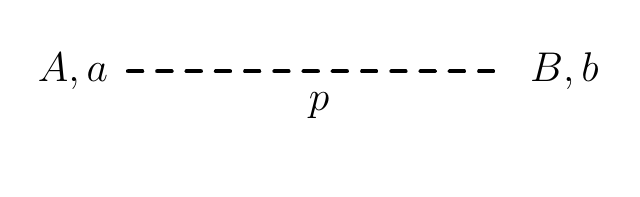}
				& \hspace{0.55cm}
				\begin{minipage}{0cm}	
					\bee
					  \frac{i \; \delta^{AB} \, \delta^{ab}}{p^2+i\epsilon}
					\eee
				\end{minipage}
		\end{tabular}
\end{table}

\subsection{Integrands}
		\label{sec:Integrand}

	All integrands are presented in four dimensions. We shall use the notation of the colour factors from \eqref{eq:ColourStructure}. The inverse of the stripped propagators is given by $D_j = Q_j^2 + i \epsilon  = q_j^2  + i\epsilon = (\sum_{k=0}^j p_k + l)^2 - \mu^2 +i \epsilon$. 
	
	Integrands which are proportional to the colour ordering $c^{adbc}$ shall be written with permuted external momenta $p_i$, i.e. $\tilde{Q_0} = l$, $\tilde{Q_1} = p_2$, $\tilde{Q_2} = l +p_2 +p_3$, $\tilde{Q_3} = Q_3$. The corresponding stripped propagators are given by $(\tilde{D}_j)^{-1} = (\tilde{Q}_j^2 + i \epsilon )^{-1}$.
	
	We only write the parts of the integrand in \YMS\, which contribute to the gravity amplitude in \theory, hence all terms that are not proportional to the colour structure \eqref{eq:ColourStructure} are not displayed. 
	
	To simplify the expression for the amplitudes we shall make use of the explicit value of the quadratic Casimir of the adjoint representation $c_A$ which for SU($N$) is given by
	\bee
		f^{a'ab'}f^{b'ba'} = c_A \delta^{ab} = N \delta^{ab}
	\eee

	\subsubsection{Integrands for $\langle 1^A_a,2^B_b,3^C_c,4^D_d\rangle$ }
		\label{sec:Integrand4scalar}

These are the two types of integrands that appear in the computation of $\left.\langle 1^A_a,2^B_b,3^C_c,4^D_d\rangle \right|_{\lambda^2 g^4}$.  The remaining integrands can be obtained by permuting the external legs. In total twelve non-equal box integrands and six non-equal triangle integrands contribute.
	\bee
		\begin{minipage}{4cm}
			\includegraphics[scale=0.6, align=c]{pics/graph1.pdf} 
		\end{minipage}
		& = & 
			\frac{\lambda^2 g^4 c^{abcd}}{D_0 D_1 D_2 D_3} F^{ABE'} F^{E'CD}
			\left[ \left(p_3 +q_3 \right) \cdot \left( q_1 -p_2 \right) - \mu^2 \right]
				\\
		\begin{minipage}{4cm}
			\includegraphics[scale=0.6, align=c]{pics/graph4.pdf} 
		\end{minipage}
		& \effequal &
		 	 - \frac{g^4 \lambda^2 c^{abdc}}{D_0 D_1 D_2} \left( F^{ADE'}F^{E'BC} + F^{ACE'}F^{E'BD} + 2 \delta^{CD} F^{A'AB'}F^{B'BA'} \right)	
	\eee
	
For the computation of $\left.\langle 1^A_a,2^B_b,3^C_c,4^D_d\rangle \right|_{g^4}$ the following diagrams have to be evaluated. $N_g=N^2-1$ is the number of adjoint generators of SU$(N)$.  By permuting the external legs the remaining non-equal integrands can be obtained (six box graphes, twelve triangles and six bubbles). 
	\bee
		\begin{minipage}{3.9cm}
			\includegraphics[scale=0.6, align=c]{pics/graph11.pdf} 
		\end{minipage}
		\hspace{-0.85cm}		
		& = & 
			\frac{g^4 c^{abcd}}{D_0 D_1 D_2 D_3} \delta^{AB} \delta^{CD}
			\left[ \left( q_1 -p_2 \right) \cdot \left(p_3 +q_3\right) - \mu^2 \right]
			\left[ \left(p_1 + q_1 \right) \cdot \left( q_3 - p_4\right) - \mu^2 \right] 
				\\
		\begin{minipage}{3.9cm}
			\includegraphics[scale=0.6, align=c]{pics/graph13.pdf} 
		\end{minipage}
		\hspace{-0.85cm}
		& \effequal & 
			\frac{g^4 c^{abdc}}{D_0 D_1 D_2} 
			\left(\delta^{AC} \delta^{BD} +\delta^{AD} \delta^{BC} -2\delta^{AB} \delta^{CD} 	\right)
		 	\left[ \left(q_0 - p_1 \right) \cdot \left( q_2 + p_2 \right) - \mu^2 \right]	
				\\
		\begin{minipage}{3.9cm}
			\includegraphics[scale=0.6, align=c]{pics/graph15.pdf} 
		\end{minipage}
		\hspace{-0.85cm}
		& \effequal &	
		 	 -2 \frac{g^4 c^{abdc}}{D_0 D_1 D_2} \delta^{AB} \delta^{CD} 	
		 	 \left[ \left(q_1-p_2 \right) \cdot \left( p_1 + q_1 \right) - \mu^2 \right]
				\\
		\begin{minipage}{3.9cm}
			\includegraphics[scale=0.6, align=c]{pics/graph17.pdf} 
		\end{minipage}
		\hspace{-0.85cm}	
		& \effequal &
		 	 2 \frac{g^4 }{D_0  D_2} c^{abdc}
		 	 \left[ 2 \left(N_g-2 \right) \delta^{AB} \delta^{CD} + \delta^{AD} \delta^{BC} + \delta^{AC} \delta^{BD} \right]
				\\
		\begin{minipage}{3.9cm}
			\includegraphics[scale=0.6, align=c]{pics/graph18.pdf} 
		\end{minipage}
		\hspace{-0.85cm}
		& \effequal &	
		 	 16 \frac{g^4}{D_0  D_2} c^{abdc} \delta^{AB} \delta^{CD}
	\eee

\subsubsection{Integrand for $\langle 1^A_a,2^B_b,3^C_c,4^+_d\rangle $}

These four types of integrands can appear in general for the computation of $\left.\langle 1^A_a,2^B_b,3^C_c,4^+_d\rangle \right|_{g^4 \lambda}$. However, once the double copy is performed the last type of graphs are vanishing. For the first type of box graphes and the triangle graphes we have two additional permutations whereas for the other box graphs we have two.

	\bee
		\begin{minipage}{3.9cm}
			\includegraphics[scale=0.6, align=c]{pics/graph21.pdf} 
		\end{minipage}
		\hspace{-0.85cm}
		 & \effequal & 
		 	i \frac{g^4 \lambda\, c^{abdc}}{D_0 D_1 D_2} F^{ABC} 	
		 	  \frac{\ketbra{r_4}{q_1 - p_2}{4}}{\sqrt{2} \, \ket{r_4}{4}}
				\\
		\begin{minipage}{3.9cm}
			\includegraphics[scale=0.6, align=c]{pics/graph22.pdf} 
		\end{minipage}
		\hspace{-0.85cm}					
		& \effequal & 0
				\\
		\begin{minipage}{3.9cm}
			\includegraphics[scale=0.6, align=c]{pics/graph19.pdf} 
		\end{minipage}
		\hspace{-0.85cm}
		& = &
			 -i \frac{g^4 \lambda\, c^{abcd}}{D_0 D_1 D_2 D_3} F^{ABC}
			\left[ \left(p_2 +q_2 \right) \cdot \left( q_0 -p_1 \right) - \mu^2 \right]
			\frac{\ketbra{r_4}{q_0 +q_3}{4}}{\sqrt{2} \, \ket{r_4}{4}}
				\\
		\begin{minipage}{3.9cm}
			\includegraphics[scale=0.6, align=c]{pics/graph20.pdf} 
		\end{minipage}
		\hspace{-0.85cm}
		& = &	
		 	 i \frac{g^4 \lambda\, c^{abcd}}{D_0 D_1 D_2 D_3} F^{ABC}
		 	\left[
		 	\left[ \left(p_1 +q_1 \right) \cdot \left( q_3 -p_4 \right) - \mu^2 \right] \frac{\ketbra{r_4}{q_2 - p_3}{4}}{\sqrt{2} \, \ket{r_4}{4}}
		 	\right.
		 			\\
		 	&& \hspace{-3cm}
		 	\left. +  
		 	\left[ 
		 		\left(q_2 - p_3 \right) \cdot \left( p_4 +q_0 \right) - \mu^2 \right] \frac{\ketbra{r_4}{p_1 +q_1 }{4}}{\sqrt{2} \, \ket{r_4}{4}}	
		 		-\left[ \left(q_2 - p_3  \right) \cdot \left( p_1 + q_1 \right) - \mu^2 \right] \frac{\ketbra{r_4}{q_0 +q_3}{4}}{\sqrt{2} \, \ket{r_4}{4}}
		 	\right]
	\eee		

\subsubsection{Integrands for $\langle 1^A_a,2^B_b,3^+_c,4^+_d\rangle$}

The gauge choice $r_3 = r_4$ reduces the amount of diagrams to compute. This gauge choice implies that only the three box graphes contribute to $\left.\langle 1^A_a,2^B_b,3^+_c,4^+_d
\rangle \right|_{g^2\lambda^2}$.
	\bee
		\begin{minipage}{3.9cm}
			\includegraphics[scale=0.6, align=c]{pics/graph24.pdf} 
		\end{minipage}
		& = &
	 - \frac{g^4 \lambda^2\, c^{abcd}}{D_0 D_1 D_2 D_3} F^{A'BB'} F^{B'AA'}
			\frac{\ketbra{r_4}{q_2+q_3}{3} \ketbra{r_4}{q_0 +q_3}{4}}{2 \, \ket{r_4}{3} \ket{r_4}{4}}
	\eee

 For $\left.\langle 1^A_a,2^B_b,3^+_c,4^+_d\rangle\right|_{g^4}$ this gauge choice sets all bubble graphs zo zero. The remaining graphs give a contribution to $\left.\langle 1^A_a,2^B_b,3^+_c,4^+_d\rangle\right|_{g^4}$. To all graphs with box topology we obtain also one other distinguished integrand. The next three triangles are the only diagrams which can be drawn with this topology. The last two graphs can be drawn in four inequivalent ways.
	\bee
		\begin{minipage}{3.9cm}
			\includegraphics[scale=0.6, align=c]{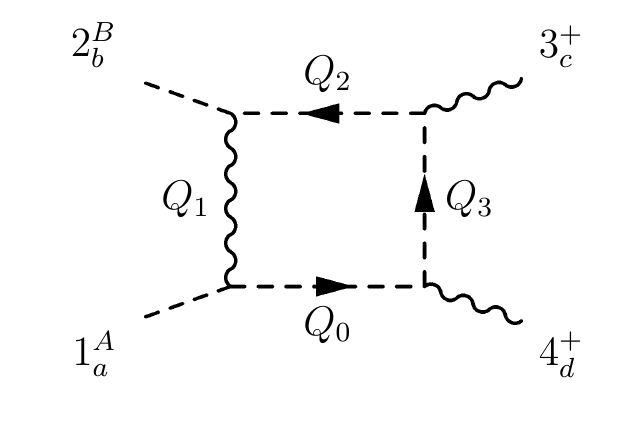} 
		\end{minipage}
		\hspace{-0.85cm}					
		& = & 
			- \frac{g^4 \, c^{abcd} \, \delta^{AB}}{D_0 D_1 D_2 D_3}
			\left(p_2 + Q_2 \right) \cdot \left( Q_0 -p_1 \right)
			\frac{\ketbra{r_4}{q_2 +q_3}{3} \ketbra{r_4}{q_0 +q_3}{4}}{2 \, \ket{r_4}{3} \ket{r_4}{4}}
				\\
		\begin{minipage}{3.9cm}		
			\includegraphics[scale=0.6, align=c]{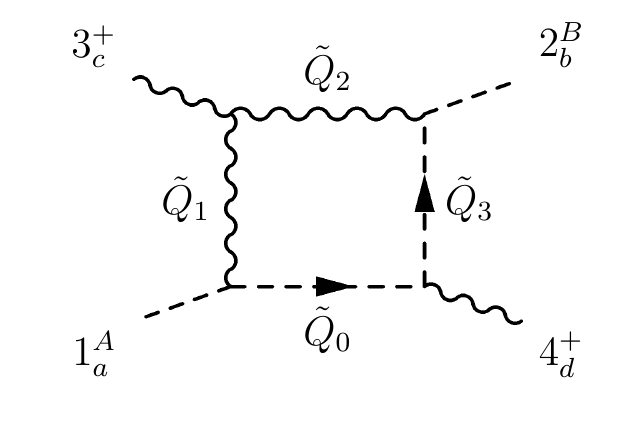} 
		\end{minipage}
		\hspace{-0.85cm}					
		& = & 
		 	 \frac{g^4 \, c^{adbc} \, \delta^{AB}}{\tilde{D}_0 \tilde{D}_1 \tilde{D}_2 \tilde{D}_3}
		 	\frac{\ketbra{r_4}{\tilde{q}_0 + \tilde{q}_3}{4}}{2 \, \ket{r_4}{3} \ket{r_4}{4}}
			\Big[
			\ketbra{r_4}{p_1 + \tilde{q}_3}{3}  \left(p_3 + \tilde{Q}_2 \right) \cdot \left( \tilde{Q}_0 -p_2 \right)					
					\\
			&& \hspace{-1cm} +
			\ketbra{r_4}{\tilde{q}_0 - p_2 }{3} \left(\tilde{Q}_1 - p_3 \right) \cdot \left( \tilde{Q}_3 + p_1 \right)
			- \ketbra{r_4}{\tilde{q}_1 + \tilde{q}_2}{3} \left(p_1 + \tilde{Q}_3 \right) \cdot \left( \tilde{Q}_0 -p_2 \right)
			\Big]
				\\
		\begin{minipage}{3.9cm}
			\includegraphics[scale=0.6, align=c]{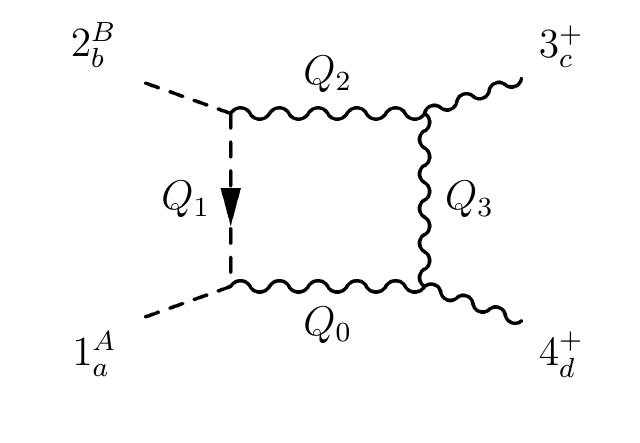} 
		\end{minipage}
		\hspace{-0.85cm}					
		& = & 	 	
		 	 -\frac{g^4 \, c^{abcd} \, \delta^{AB}}{2\, D_0 D_1 D_2 D_3}
		 	 \Big[
		 	 \frac{\ketbra{r_4}{q_1 - p_2}{3} \ketbra{r_4}{q_2 - p_3}{4} - \ketbra{r_4}{q_3 + q_2}{3} \ketbra{r_4}{q_1 - p_2}{4}}{\ket{r_4}{3} \ket{r_4}{4}}
		 	 		\\
		 	 && \hspace{-1cm} \times 
		 	 \left(Q_3 - p_4  \right) \cdot \left( p_1 + Q_1 \right)
		 	 +\frac{\ketbra{r_4}{p_1 + q_1}{4}}{\ket{r_4}{3} \ket{r_4}{4}}
		 	 \big[
		 	 \ketbra{r_4}{p_4 + q_0}{3} \left(Q_1 -p_2\right) \cdot \left(Q_3 +p_3 \right)
		 	 		\\
		 	 && \hspace{-1cm}
		 	 +\ketbra{r_4}{q_1 - p_2}{3} \left(Q_2 -p_3\right) \cdot \left(Q_0 +p_4 \right)
		 	 -\ketbra{r_4}{q_2 + q_3}{3} \left(Q_1 -p_2\right) \cdot \left(Q_0 +p_4 \right)
		 	 \big]
		 	 		\\
		 	 && \hspace{-1cm}
		 	 -\frac{\ketbra{r_4}{q_0 + q_3}{4}}{\ket{r_4}{3} \ket{r_4}{4}}
		 	 \big[
		 	 \ketbra{r_4}{q_1 - p_2}{3} \left(Q_2 -p_3\right) \cdot \left(Q_1 +p_1 \right)
					\\
		 	  && \hspace{-1cm}
		 	  +\ketbra{r_4}{p_1 + q_1}{3} \left(Q_1 -p_2\right) \cdot \left(Q_3 +p_3 \right)
		 	  -\ketbra{r_4}{q_2 + q_3}{3} \left(Q_1 -p_2\right) \cdot \left(Q_1 +p_1 \right)
		 	 \big]
		 	 \Big]
						\\
		\begin{minipage}{3.9cm}
			\includegraphics[scale=0.6, align=c]{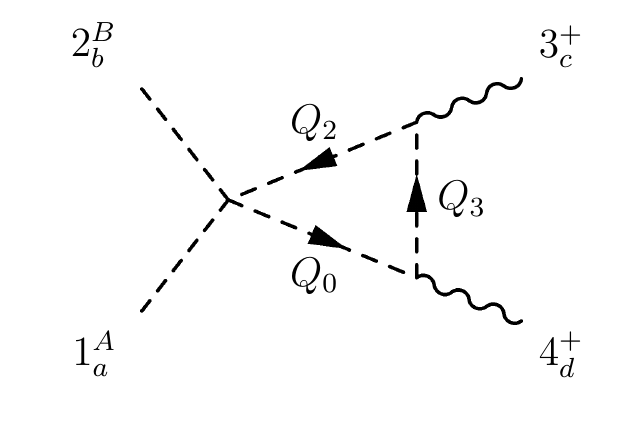} 
		\end{minipage}
		\hspace{-0.85cm}					
		& \effequal & 
		 	  - \frac{g^4 \, c^{abcd} }{ D_0 D_2 D_3} \delta^{AB}
		 	  \frac{\ketbra{r_4}{q_2 +q_3}{3} \ketbra{r_4}{q_0 +q_3}{4}}{ \ket{r_4}{3} \ket{r_4}{4}}
		 	\left( 1 - N_g \right)
						\\
		\begin{minipage}{3.9cm}
			\includegraphics[scale=0.6, align=c]{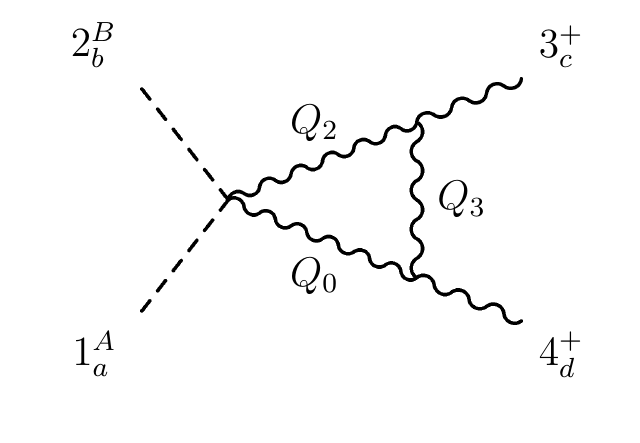} 
		\end{minipage}
		\hspace{-0.85cm}					
		& \effequal & 
		 	  \frac{g^4 \, c^{abcd} }{ D_0 D_2 D_3} \frac{\delta^{AB}}{ \ket{r_4}{3} \ket{r_4}{4}}
		 	 \Big[
		 	 2 \ketbra{r_4}{q_2 +q_3}{3} \ketbra{r_4}{q_0 +q_3}{4}
		 	 		\\
		 	&&  \hspace{-1cm}
		 	 +\ketbra{r_4}{q_0 +p_4}{3} \ketbra{r_4}{p_3 +q_3}{4}
		 	+\ketbra{r_4}{q_3 - p_4}{3} \ketbra{r_4}{q_2 - p_3}{4}
		 	\Big]
					\\
		\begin{minipage}{3.9cm}
			\includegraphics[scale=0.6, align=c]{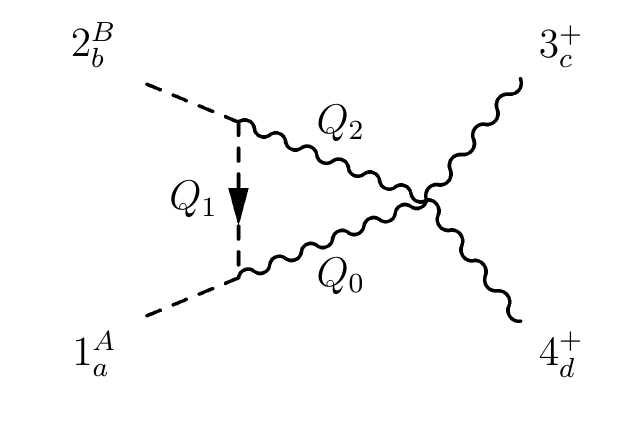} 
		\end{minipage}
		\hspace{-0.85cm}					
		& \effequal & 
		 	 - \frac{g^4 \, c^{abcd} }{ D_0 D_1 D_2}  \delta^{AB} 
		 	 \left[
		 	 \frac{\ketbra{r_4}{q_1 - p_2}{3} \ketbra{r_4}{p_1 +q_1}{4}}{2\, \ket{r_4}{3} \ket{r_4}{4}}
		 	 +\frac{\ketbra{r_4}{p_1 + q_1}{3} \ketbra{r_4}{q_1 - p_2}{4}}{2\, \ket{r_4}{3} \ket{r_4}{4}}
		 	 \right]
		 	 	\\
		\begin{minipage}{3.9cm}
			\includegraphics[scale=0.6, align=c]{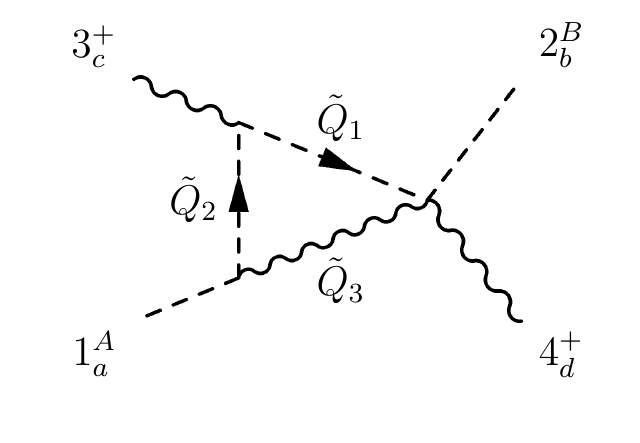}			
		\end{minipage}
		\hspace{-0.85cm}					
		& \effequal & 
		 	- \frac{g^4 \, c^{adbc} }{\tilde{D_1} \tilde{D}_2 \tilde{D}_3}  \delta^{AB} 
		 	 \frac{\ketbra{r_4}{\tilde{q}_1 + \tilde{q}_2}{3} \ketbra{r_4}{p_1 - \tilde{q}_2}{4}}{2\, \ket{r_4}{3} \ket{r_4}{4}}	
					\\
		\begin{minipage}{3.9cm}
			\includegraphics[scale=0.6, align=c]{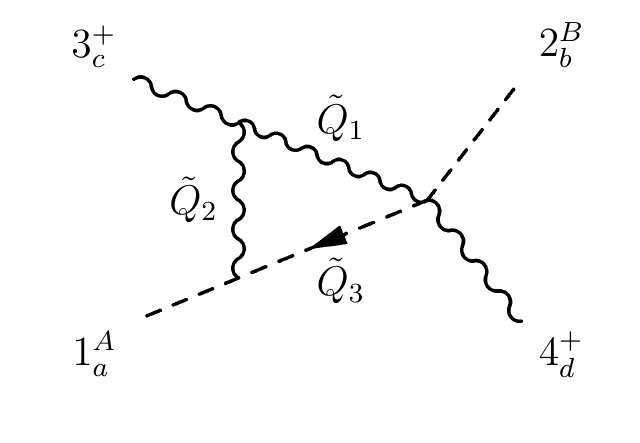}			
		\end{minipage}
		\hspace{-0.85cm}					
		& \effequal & 
		 	- \frac{g^4 \, c^{adbc} }{ \tilde{D_1} \tilde{D}_2 \tilde{D}_3}  \delta^{AB}
		 	\Big[ 
		 	 \frac{\ketbra{r_4}{\tilde{q}_3 + p_1}{3} \ketbra{r_4}{p_3 + \tilde{q}_2}{4}}{2\ket{r_4}{3} \ket{r_4}{4}}
		 	  - \frac{\ketbra{r_4}{\tilde{q}_1 + \tilde{q}_2}{3} \ketbra{r_4}{p_1 + \tilde{q}_3}{4}}{2\ket{r_4}{3} \ket{r_4}{4}}
		 	 \Big]
		\eee

\newpage

\bibliographystyle{nb}
\bibliography{bibliothek}

\end{document}